\documentstyle[aps,preprint,psfig,eqsecnum]{revtex}
\newcommand{\half}{{1 \over 2}}
\newcommand{\quarter}{{1 \over 4}}

\tightenlines
\begin{document}
\draft
\def\nva{v_1 \cdot n}
\def\nvb{v_2 \cdot n}
\def\v1v2{v_1 \cdot v_2}
\title{Post-Newtonian gravitational radiation
and equations of motion via direct
integration of the relaxed Einstein equations. \\
II. Two-body equations of motion to second post-Newtonian 
order, and radiation-reaction to 3.5 post-Newtonian order
}
\author{Michael E. Pati and Clifford M. Will }
\address{McDonnell Center for the Space Sciences,
Department of Physics, \\
Washington University, St. Louis, Missouri 63130}
\date{\today}
\maketitle
\begin{abstract}

We derive the equations of motion for binary systems of compact bodies in
the post-Newtonian (PN) approximation to general relativity.  Results are given
through 2PN order (order $(v/c)^4$ beyond Newtonian theory), and for 
gravitational radiation reaction effects at 2.5PN and 3.5PN orders.  
The method is based
on a framework for direct integration of the relaxed Einstein equations
(DIRE) developed earlier, in which the equations of motion through 3.5PN
order can be expressed
in terms of Poisson-like potentials that are generalizations of the
instantaneous Newtonian gravitational 
potential, and in terms of multipole moments of the system and their time
derivatives.  All potentials are well defined and free of divergences
associated with integrating quantities over all space.  Using a model of the
bodies as spherical, non-rotating fluid balls whose characteristic size $s$
is small compared to the bodies' separation $r$, we develop a method for
carefully extracting only terms that are independent of the parameter $s$,
thereby ignoring tidal interactions, spin effects, and internal self-gravity
effects.  Through 2.5PN order, the resulting equations agree completely 
with those
obtained by other methods; the new 3.5PN back-reaction
results are shown to be consistent
with the loss of energy and angular momentum via radiation to infinity.  

\end{abstract}
%
\section{Introduction and Summary}
\label{sec:intro}

This is the second in a series of papers which will treat
motion and gravitational radiation in the post-Newtonian
approximation to general relativity.  While this is a problem that dates
back to the beginnings of general relativity, it has recently taken on added
observational importance because of the need for extremely accurate
theoretical gravitational waveform templates for analysis of
data taken by laser interferometric gravitational-wave detectors
\cite{3min}.  
Specifically, for waves from inspiralling binary systems of compact objects
(neutron stars or black holes), equations of motion and gravitational
waveforms accurate
to at least {\it third} post-Newtonian order (order $(v/c)^6$) beyond the
initial Newtonian or quadrupole approximation are needed.  

In paper I \cite{patiwill}, we laid out the foundations of our method
of Direct Integration of the Relaxed Einstein Equations (DIRE).  
We rewrote the Einstein equations  as a flat spacetime wave equation
together with a harmonic gauge condition (the ``relaxed'' Einstein
equations), and solved them formally in terms
of a retarded integral over the past null cone of the field point.  Because
the ``source'' contains both the material stress-energy tensor and the
stress-energy contributions of the 
gravitational fields themselves, it was necessary to iterate the integrals
repeatedly to obtain successively higher-order approximations 
to a solution in powers of
$\epsilon \sim (v/c)^2 \sim (Gm/rc^2)$.  Each power of 
$\epsilon$ represents one
``post-Newtonian'' (PN) order in the series ($\epsilon^{1/2}$ represents
one half, or 0.5PN orders).  Despite the fact that the field
contributions to the integrals extend over all spacetime, we demonstrated
that no infinite or ill-defined integrals occurred, even in
slow-motion, multipole expansions, and found a simple
prescription for evaluating the finite contributions of all integrals.  
This was true for calculations of the metric both in the near zone and in
the far zone.

To complete the solution of Einstein's equations, one needs 
equations of motion for the system.  For this, one needs the spacetime
metric evaluated for field points within the near zone, corresponding to a
sphere of radius $\cal R \sim$ one gravitational wavelength.  In Paper I, we
expressed this near-zone metric explicitly through order 
$\epsilon^{7/2}$ beyond
the Newtonian approximation, corresponding to 3.5 post-Newtonian (PN) order,
in terms of instantaneous, Poisson-like integrals and their
generalizations, of the form, for example,
\begin{equation}
P(f) \equiv {1 \over {4\pi}} \int_{\cal M} {{f(t,{\bf x}^\prime)}
\over {|{\bf x}-{\bf x}^\prime | }} d^3x^\prime \,,
\label{poisson1}
\end{equation}
where the integration is confined to the near zone $\cal M$, and only that
part of the integral that is independent of $\cal R$ is kept.

It is the purpose of this paper to evaluate these
integrals explicitly for a binary system of non-spinning, spherically
symmetric bodies whose size is much  smaller than their separation.  We
will carry this evaluation through 2PN order, and will also evaluate the
leading radiation-reaction contributions at 2.5PN order, together with the
first post-Newtonian corrections to radiation reaction, at 3.5PN order.
The extremely lengthy derivation of the 
non-radiative 3PN contributions will be reserved
for future publications.  The resulting equations have
the form
\begin{eqnarray}
{{d^2 {\bf x}} \over {dt^2}} &=& -{m \over r^2} {\bf n} 
+ {m \over r^2} \bigl[ \, {\bf n} (A_{PN} + A_{2PN} + A_{3PN}) 
	+ {\dot r}{\bf v} (B_{PN} + B_{2PN} + B_{3PN}) \bigr]  
\nonumber \\
&&
+ {8 \over 5} \eta {m \over r^2} {m \over r} 
	\bigl[\dot r {\bf n} (A_{2.5PN}+A_{3.5PN})
	- {\bf v}(B_{2.5PN}+B_{3.5PN})\bigr] \,,
\label{eomfinal}
\end{eqnarray}
where ${\bf x} \equiv {\bf x}_1 - {\bf x}_2$, $r \equiv |{\bf x}|$, ${\bf n}
\equiv {\bf x}/r$, $m \equiv m_1 + m_2$, $\eta \equiv m_1m_2/m^2$, 
${\bf v} \equiv {\bf v}_1 - {\bf v}_2$, and $\dot
r = dr/dt$.  
We use units in which $G = c = 1 $.  The leading term is Newtonian gravity.
The other terms on the first line are the ``conservative'' or
non-dissipative terms, of even PN order, while those on the second line are
dissipative radiation-reaction terms, of odd-half PN order.  The
coefficients $A$ and $B$ are given explicitly by
\begin{mathletters}
\begin{eqnarray}
A_{PN} &=& -(1+3\eta)v^2 + {3 \over 2}\eta {\dot r}^2 +2(2+\eta)m/r \,,
\nonumber \\
B_{PN} &=&  2(2-\eta) \,, \label{eomfinalcoeffsPN}
\\
A_{2PN} &=& -\eta(3-4\eta)v^4 + {1 \over 2}\eta(13-4\eta)v^2m/r
	+{3 \over 2}\eta(3-4\eta)v^2{\dot r}^2 
\nonumber \\
&&
	+(2 +25\eta+2\eta^2){\dot r}^2m/r
	-{15 \over 8}\eta(1-3\eta){\dot r}^4
	-{3 \over 4}(12+29\eta)(m/r)^2 \,,
\nonumber \\
B_{2PN} &=& {1 \over 2}\eta(15+4\eta)v^2
	-{3 \over 2}\eta(3+2\eta){\dot r}^2
	-{1 \over 2}(4 +41\eta+8\eta^2)m/r \,,
\label{eomfinalcoeffs2PN}
\\
A_{2.5PN} &=& 3v^2+{17 \over 3}m/r\,,
\nonumber \\
B_{2.5PN} &=& v^2 + 3m/r\,,
\label{eomfinalcoeffs2.5PN}
\\
A_{3.5PN} &=&
-{3 \over 28} (61 + 70\eta) v^4
-{1 \over 42} (519 - 1267\eta) v^2m/r
+{15 \over 4}(19 + 2\eta) v^2\dot r^2
\nonumber \\
&&
-{1 \over 4}(147 + 188\eta) \dot r^2m/r
-70 \dot r^4
-{23 \over 14}(43 +14\eta) (m/r)^2 \,,
\nonumber \\
B_{3.5PN} &=&
-{1 \over 28}(313 + 42\eta )v^4
+{1 \over 42}(205 +777\eta) v^2m/r
+{3 \over 4}(113 +2 \eta) v^2\dot r^2
\nonumber \\
&&
-{1 \over 12}(205 + 424 \eta) \dot r^2m/r
-75\dot r^4
-{1 \over 42}(1325 + 546\eta) (m/r)^2 \,.
\label{eomfinalcoeffs3.5PN}
\end{eqnarray}
\label{eomfinalcoeffs}
\end{mathletters}

The 1PN coefficients are standard; the 2PN and 2.5PN 
terms agree completely with results derived
by Damour and Deruelle \cite{DD81,damour300}, 
Kopeiken and Grishchuk \cite{kopeikin85,GK86}, Blanchet
{\it et al.} \cite{bfp98}, and Itoh {\it et al.} \cite{futamase01}.  
However, the 2.5PN terms
differ from those derived, for example from a Burke-Thorne type 
radiation-reaction potential given by $\Phi_{RR} = -(1/5)x^ix^j 
d^5 {\cal I}^{<ij>}/dt^5$, where ${\cal I}^{<ij>}$ is the system's
trace-free quadrupole moment (see, e.g. \S 36.11 of \cite{MTW}). 
Iyer and Will
\cite{iyerwill,iyerwill2}
showed that there is a two-parameter ``gauge'' freedom in the radiation
reaction equations at 2.5PN order (and a six-parameter freedom at 3.5 PN
order), within which the different equations of motion yield identical
results for the net energy and angular momentum radiated.  The 2.5PN terms
shown are thus observationally equivalent to the Burke-Thorne radiation
reaction equations.  The 3.5PN terms are new; it can be shown that they
correspond to a specific choice of the six Iyer-Will gauge parameters, and
thus automatically generate the proper post-Newtonian corrections to energy
and angular momentum loss.  Sch\"afer and Jaranowski 
\cite{schafer85,schafer86,jaraschafer97}
also derived 2PN, 2.5PN and 
3.5PN contributions to the equations of motion in a Hamiltonian
formulation.   The 3PN contributions to the equations of motion have
also been reported by several groups 
\cite{jaraschafer98,jaraschafer99,djs00,bf00,bf01}.

The remainder of this paper is devoted to the details supporting these
results.  In Sec. \ref{sec:eombasic} 
we review the basic equations needed to find
equations of motion to the order needed.  
Section \ref{sec:2body}
specializes to binary systems of 
spherical ``pointlike'' bodies, and derives the equations of
motion for each body correct to 2PN order.  Section \ref{sec:2bodyreaction} 
repeats the
process for the 2.5PN and 3.5PN terms.  In Sec. \ref{sec:relative} 
we transform to an
effective one-body relative equation of motion.  Concluding remarks are made
in Sec. \ref{sec:conclusions}.  Detailed points are
reserved for a series of Appendices.

Our conventions and notation generally follow those of
\cite{MTW,thorne80}.  
Greek indices run over four spacetime values 0, 1, 2, 3, while
Latin indices run over three spatial values 1, 2, 3;
commas denote partial derivatives with
respect to a chosen coordinate system, while semicolons denote
covariant derivatives;
repeated indices are summed over;
$\eta^{ \mu \nu } = \eta_{ \mu \nu } = {\rm diag}(-1,1,1,1)$;
$g \equiv \det(  g_{ \mu \nu } )$;
$a^{(ij)} \equiv ( a^{ij} + a^{ji} )/2$;
$a^{[ij]} \equiv ( a^{ij} - a^{ji} )/2$;
$\epsilon^{ijk}$ is the totally antisymmetric Levi-Civita symbol
$( \epsilon^{123} = + 1)$.  We use a multi-index notation for products
of vector components and partial derivatives, and for multiple spatial
indices:
$x^{ij \dots k} \equiv x^ix^j \dots x^k$,
$\nabla^{ij \dots k} \equiv \nabla^i\nabla^j \dots \nabla^k$, with a
capital letter superscript denoting an abstract product of that
dimensionality:
$x^Q \equiv x^{ i_1 } x^{ i_2 } ... x^{ i_q } $ and
$\nabla^Q \equiv \nabla^{ i_1 } \nabla^{ i_2 } ... \nabla^{ i_q } $.
Also,  for a tensor of rank $Q$, $f^Q \equiv f^{i_1 i_2 \dots i_q}$.
The notation $f^Q g^Q$ denotes a complete contraction over the $q$
indices.
Spatial indices are freely raised and lowered with
$\delta^{ij}$ and $\delta_{ij}$.

\section{Equations of motion of compact binary systems: basic
equations}
\label{sec:eombasic}

\subsection{Structure of the near-zone metric to 3.5PN order}
\label{sec:nearzone}

We begin by reviewing key results from Paper I.  We defined the ``field''
$h^{\alpha \beta}$ by
\begin{eqnarray}
h^{\alpha \beta} \equiv \eta^{\alpha \beta} - (-g)^{1/2} g^{\alpha \beta} \;
,
\label{hdefinition}
\end{eqnarray}
where $g^{\alpha \beta}$ is the space time metric.  In deDonder
or harmonic coordinates defined by the gauge condition
$h^{\alpha \beta},_{\beta} = 0$,  
the Einstein equations take the
form
\begin{eqnarray}
\Box h^{ \alpha \beta } = -16 \pi {\tau}^{ \alpha \beta } \; ,
\label{relaxed}
\end{eqnarray}
where $\Box $
is the flat-spacetime wave operator, and ${\tau}^{ \alpha \beta }$ is made
up of the material stress-energy tensor $T^{ \alpha \beta }$ 
and the contribution of all the
non-linear terms in Einstein's
equations.
We defined a notation for specific components of the field
$h^{\alpha\beta}$:
\begin{eqnarray}
N &\equiv& h^{00} \sim O(\epsilon) \,, \nonumber \\
K^i &\equiv& h^{0i} \sim O(\epsilon^{3/2}) \,, \nonumber \\
B^{ij} &\equiv& h^{ij} \sim O(\epsilon^2) \,, \nonumber \\
B &\equiv& \sum_i h^{ii} \sim O(\epsilon^2) \,, 
\label{hcomponents}
\end{eqnarray}
where we show the leading-order dependence on $\epsilon$ in the near
zone.  To the necessary orders for calculating
the equations of motion to 3.5PN order, 
the components of the physical metric 
are given in terms of $N$, $K^i$, $B^{ij}$ and $B$ by
\begin{eqnarray}
g_{00} &=& -(1- {1 \over 2}N+ {3 \over 8} N^2 - {5 \over 16} N^3 + {35
\over 128}N^4 ) + {1 \over 2}B(1- {1 \over 2}N+ {3 \over 8} N^2)
\nonumber \\
&&+ {1 \over 4}(B^{ij}B^{ij}-{1 \over 2}B^2) 
+{1 \over 2}K^jK^j -  {3 \over 4} NK^jK^j +O(\epsilon^5) \,, 
\nonumber \\
g_{0i} &=& -K^i(1- {1 \over 2}N -{1 \over 2}B + {3 \over 8} N^2) - K^j
B^{ij} +O(\epsilon^{9/2}) \,, 
\nonumber \\
g_{ij} &=& \delta^{ij} (1+ {1 \over 2}N- {1 \over 8} N^2 + {1 \over 16}
N^3 - {1 \over 4}NB +{1 \over 2}K^kK^k ) 
\nonumber \\
&&+ B^{ij} - {1 \over
2}B\delta^{ij}- K^iK^j + {1 \over 2}N B^{ij} +O(\epsilon^4) \,, 
\nonumber \\
(-g) &=& 1+N-B-NB+K^iK^i + O(\epsilon^4) \,.
\label{metricexpand}
\end{eqnarray}
The potentials $N$, $K^i$,
$B^{ij}$ and $B$ must also be expanded to an appropriate order:
\begin{eqnarray}
N &=& \epsilon (4U_{\sigma}  
+ \epsilon N_1+ \epsilon^{3/2} N_{1.5}+ \epsilon^2 N_2+
\epsilon^{5/2} N_{2.5}+ \epsilon^3 N_3+ \epsilon^{7/2} N_{3.5} )
+O(\epsilon^5) \,, 
\nonumber \\
K^i &=& \epsilon^{3/2} (4V_{\sigma}^i + \epsilon K_2^i +\epsilon^{3/2} K_{2.5}^i
+ \epsilon^2 K_3^i +\epsilon^{5/2} K_{3.5}^i ) +O(\epsilon^{9/2}) \,, 
\nonumber \\
B &=& \epsilon^2 (B_1 + \epsilon^{1/2} B_{1.5}
+\epsilon B_2 + \epsilon^{3/2} B_{2.5} + \epsilon^2 B_3+
\epsilon^{5/2} B_{3.5} )
+O(\epsilon^5) \,, 
\nonumber \\
B^{ij} &=& \epsilon^2 (B_2^{ij} + \epsilon^{1/2} B_{2.5}^{ij} +
\epsilon B_3^{ij}+ \epsilon^{3/2} B_{3.5}^{ij}) +O(\epsilon^4) \,,
\label{expandNKB}
\end{eqnarray}
where the subscript on each term indicates the level (1PN, 2PN, 2.5PN,
etc.) of its leading contribution to the equations of motion.  Notice
that our separate treatment of $B$ and $B^{ij}$ leads to the slightly
awkward notational circumstance that, for example, $B_2^{ii} = B_1$.
In Paper I, we obtained 
explicit near-zone expressions for each of the terms 
in Eq. (\ref{expandNKB}) 
in terms of Poisson-like potentials
\begin{eqnarray}
P(f) &\equiv& {1 \over {4\pi}} \int_{\cal M} {{f(t,{\bf x}^\prime)}
\over {|{\bf x}-{\bf x}^\prime | }} d^3x^\prime \,, \quad \nabla^2
P(f) = -f \,, 
\nonumber\\
\Sigma (f) &\equiv& \int_{\cal M} {{\sigma(t,{\bf x}^\prime)f(t,{\bf
x}^\prime)}
\over {|{\bf x}-{\bf x}^\prime | }} d^3x^\prime = P(4\pi\sigma f) \,,
\label{poissonlike}
\end{eqnarray}
and various generalizations [for definitions, see Paper I, Eqs. (4.10) --
(4.16)],
and in terms of source multipole moments, such as 
${\cal I}^{ij} \equiv \int_{\cal M} \tau^{00} x^{ij} d^3x$ [for definitions, 
see Paper I, Eqs. (2.14) and (4.5) -- (4.7)].  
The integrations are over a constant time hypersurface $\cal M$ that extends
to a radius ${\cal R} \sim$ one gravitational wavelength from the source.
In Paper I, we showed that, even for Poisson potentials where the function
$f$ does not have compact support, all contributions to the field
$h^{\mu\nu}$ from the integration
over $\cal M$ that depend on $\cal R$ cancel corresponding contributions
from that part of the field point's past null cone that is outside $\cal
R$, and thus that any $\cal R$-dependent terms that appear in a 
given integral can simply be discarded.

The potentials given in Paper I 
were expressed in terms of specific source 
densities given by
\begin{eqnarray}
\sigma & \equiv& T^{00} + T^{ii} \,, \nonumber \\
\sigma^i & \equiv& T^{0i} \,, \nonumber \\
\sigma^{ij} & \equiv& T^{ij} \,.
\label{sigmadefinitions}
\end{eqnarray}
For example, in Eq. (\ref{expandNKB}), 
\begin{eqnarray}
U_{\sigma} &\equiv& \int_{\cal M} {{\sigma(t,{\bf x}^\prime)}
\over {|{\bf x}-{\bf x}^\prime | }} d^3x^\prime = P(4\pi\sigma) =
\Sigma(1) \,, 
\nonumber \\
V_{\sigma}^i &\equiv& \int_{\cal M} {{\sigma^i(t,{\bf x}^\prime)}
\over {|{\bf x}-{\bf x}^\prime | }} d^3x^\prime = P(4\pi\sigma^i) =
\Sigma^i(1) \,.
\label{UVsigma}
\end{eqnarray}
Explicit expressions for the remaining terms in Eq. (\ref{expandNKB}) can
be found in Paper I, Eqs. (5.2), (5.4), (5.8), (5.10), (6.2), and (6.4).

\subsection{Model of the material sources}
\label{sourcemodel}

We model the material sources in the binary system as
perfect fluid, having
stress-energy tensor
\begin{eqnarray}
T^{\alpha\beta} \equiv (\rho +p)u^\alpha u^\beta +pg^{\alpha\beta}
\;,
\label{fluid}
\end{eqnarray}
where $\rho$ and $p$ are the locally measured energy density and
pressure, respectively, and $u^\alpha$ is the four-velocity of an
element of fluid.
We will assume the bodies to be non-rotating (the effects of spin will be
treated in future publications), spherically symmetric in their comoving rest
frames, and small compared to their separation, so that tidal
distortions can be ignored.  

Our goal is to determine all
contributions to the equations of motion that are independent of the
internal structure, size, and shape of the bodies.   
We are less interested in formal rigor than in having a robust method
that captures all the effects without missing any.
One approach to this has been to assume a ``delta-function'' or
distributional form for the stress-energy tensor.  This has been
criticized because such a source is fundamentally
incompatible with general
relativity, and because it leads to divergences related to the infinite
self-field of a point mass.  A number of methods have been developed
in order to extract the finite part of such divergent expressions,
including the Hadamard {\it partie finie} technique (for a recent
review, see \cite{bfhadamard}.
Another approach is related to that of
Einstein, Infeld and Hoffman (EIH) \cite{EIH}: expand the vacuum
Einstein equations in a post-Newtonian expansion and match the
solutions to fields representing the near-zone, Schwarzschild-like
field of a static, spherical body.  The consistency conditions imposed
by the matching lead to constraints on the motion of the bodies that
yield the equations of motion.  This has been carried out to 2.5PN
order by Itoh {\it et al.} \cite{futamase00,futamase01}, exploiting a
``strong-field point particle'' scaling method of Futamase
\cite{futamase85}.

A third approach is to treat the bodies realistically as fluid balls
with internal energy, supported against their self gravity by pressure
governed by an equation of state.  In this case, the mass of each body
is composed of rest mass, internal energy and self-gravitational
binding energy, and the center of mass is defined accordingly.
However, at Newtonian and 1PN order, when finite-size effects such as tidal
interactions are ignored, it turns out that all vestiges of the internal
structure are ``effaced'', in the language of Damour \cite{damour300},
and the final 1PN equations of motion depend on one and only one mass
as defined above.  This procedure can be seen in detail, for example,
in \cite{tegp}, \S 6.2, where the calculation is actually carried
out in the parametrized post-Newtonian (PPN) framework, which
encompasses a class of metric theories of gravity.  In many
alternative theories, such as scalar-tensor gravity, the effacement is
violated, and the equations of motion depend on various masses, such
as inertial mass $m$, active gravitational mass $m_A$, and passive
gravitational mass $m_P$, which may differ by amounts depending on
the bodies'
gravitational binding energy.  In GR, the PPN parameters are such that
all three masses are identical.  

In fact, the 1PN equations of motion derived from this method are
identical to those obtained from a ``delta'' function method in which
one systematically throws away all terms that are singular when
evaluated on each body's world line.  At 1PN order,
the results are in keeping with
the idea that general relativity satisfies the Strong Equivalence
Principle (see \S 3.3 of \cite{tegp}), part of which implies that the
motion of bound bodies is independent of their internal structure,
provided that tidal effects can be ignored.  
Kopeikin \cite{kopeikin85} extended this to 2PN
order, with results consistent with the Strong Equivalence Principle.  

Our approach will be intermediate between the ``delta-function'' model and
the full equilibrium fluid ball method.  We will neglect pressure $p$ and
internal energy density, and treat the bodies as balls of baryons
characterized by the ``conserved'' baryon mass density $\rho^*$, given by
\begin{equation}
\rho^* \equiv m n \sqrt{-g} u^0 \,,
\label{conservedrho}
\end{equation}
where $m$ is the rest mass per baryon, $n$ is the baryon number
density, and 
$g \equiv \det (g_{\mu\nu})$.  From the conservation of baryon number,
expressed in covariant terms by $(nu^{\alpha})_{;\alpha} = 0 =
(\sqrt{-g}nu^{\alpha})_{,\alpha}$, we see that $\rho^*$ obeys the non-covariant,
but exact, continuity
equation
\begin{equation}
\partial \rho^* /\partial t + \nabla \cdot (\rho^* {\bf v}) = 0 \,,
\label{continuity}
\end{equation}
where $v^i = u^i/u^0$, and spatial gradients and dot products use a
Cartesian metric.
In terms of $\rho^*$, the stress-energy tensor takes the form
\begin{equation}
T^{\alpha\beta} = \rho^* (-g)^{-1/2} u^0 v^\alpha v^\beta \,,
\label{Trhostar}
\end{equation} 
where $v^\alpha = (1,v^i)$.  We define the baryon rest mass, center of baryonic
mass, velocity
and acceleration of each body by the formulae
\begin{eqnarray}
m_A &\equiv & \int_A \rho^* d^3x \,, 
\nonumber \\
{\bf x}_A & \equiv & (1/m_A) \int_A \rho^* {\bf x}^i d^3x \,, 
\nonumber \\
{\bf v}_A & \equiv & d{\bf x}_A /dt = (1/m_A) \int_A \rho^* {\bf v}^i d^3x \,, 
\nonumber \\
{\bf a}_A & \equiv & d{\bf v}_A /dt = (1/m_A) \int_A \rho^* {\bf a}^i d^3x \,, 
\label{rhostardefinitions}
\end{eqnarray}
where we have used the general fact, implied by the equation of continuity for
$\rho^*$, that 
\begin{equation}
{\partial \over {\partial t}} \int \rho^*(t, {\bf x}^\prime) f(t, {\bf x},{\bf
x}^\prime) d^3x^\prime = \int \rho^*(t, {\bf x}^\prime) \left ( {\partial
\over {\partial t}} + {\bf v}^\prime \cdot \nabla^\prime \right ) f(t, {\bf
x},{\bf x}^\prime) d^3x^\prime \,.
\label{continuity2}
\end{equation}

\subsection{Structure of the equations of motion to 3.5 PN order}

The definition of the stress-energy tensor in terms of $\rho^*$, Eq.
(\ref{Trhostar}), together with the equation of continuity, Eq.
(\ref{continuity}), and the fundamental equations of motion,
${T^{\alpha\beta}}_{;\beta}=0$ can be shown to be equivalent to the
geodesic equation $u^{\beta}u^\alpha_{;\beta} =0$ for each fluid
element.  In terms of ordinary velocity $v^i = dx^i/dt$ and harmonic coordinate
time $t$, the geodesic equation takes the form
\begin{equation}
a^i \equiv dv^i/dt = -\Gamma^i_{\alpha\beta} v^\alpha v^\beta +
\Gamma^0_{\alpha\beta} v^\alpha v^\beta v^i \,,
\label{geodesic}
\end{equation}
where $\Gamma^\gamma_{\alpha\beta}$ are Christoffel symbols computed
from the metric.  According to our definitions of the baryonic center
of mass, velocity and acceleration of each body, we can write the
coordinate acceleration  of the $A$-th body in the form
\begin{equation}
a^i_A = (1/m_A) \int_A \rho^* (-\Gamma^i_{\alpha\beta} v^\alpha
v^\beta +
\Gamma^0_{\alpha\beta} v^\alpha v^\beta v^i ) d^3x \,.
\label{acceleration1}
\end{equation}
Our task therefore, is to determine the Christoffel symbols through a
PN order sufficient for equations of motion valid through 3.5PN order
using the 3.5PN accurate expressions of the metric in  Paper I
(different components of $\Gamma^\gamma_{\alpha\beta}$ are need to
different accuracy, depending on the number of factors of velocity
which multiply them); re-express the Poisson potentials contained in
the metric in terms of $\rho^*$, rather than in terms of the
``densities'' $\sigma$, $\sigma^i$ and $\sigma^{ij}$, substitute into
Eq. (\ref{acceleration1}), and integrate over the $A$-th body, keeping
only terms that do not depend on the bodies' finite size.

\subsection{Christoffel Symbols to 3.5PN order}

The fundamental definition
\begin{equation}
\Gamma^{\alpha}_{\mu\nu} \equiv {1 \over 2} g^{\alpha\beta} 
	(g_{\beta\mu , \nu}+ g_{\beta\nu , \mu} - g_{\mu\nu , \beta}) \,,
\end{equation}
together with the form of the metric Eq. (\ref{metricexpand}) and the
expansions of Eq. (\ref{expandNKB}) give the Christoffel
symbols expanded to the required order:
\begin{mathletters}
\begin{eqnarray}
\Gamma^0_{00} &=&
-\epsilon \dot U_{\sigma}
-\epsilon^2 \left ( \quarter ({\dot N}_1 + {\dot B}_1) - 
	4U_{\sigma} \dot U_{\sigma} 
- 4 V_{\sigma}^i U_{\sigma}^{,i} 
	\right ) 
	-\epsilon^{5/2} {\dot N}_{1.5}
	-\epsilon^3 \biggl (\quarter ({\dot N}_2 + 
	{\dot B}_2) -N_1 \dot U_{\sigma} 
\nonumber \\
&&
- U_{\sigma} \dot N_1
	- V_{\sigma}^i (N_1^{,i} + B_1^{,i} ) 
	- K_2^i U_{\sigma}^{,i} 
	- 8 V_{\sigma}^i \dot V_{\sigma}^i
	+ 16 U_{\sigma}^2 \dot U_{\sigma} 
	+32 U_{\sigma} V_{\sigma}^i U_{\sigma}^{,i} \biggr ) 
\nonumber \\
&&
- \epsilon^{7/2} \left ( \quarter (\dot N_{2.5} + \dot B_{2.5})
	-N_{1.5} \dot U_{\sigma} 
	- U_{\sigma} \dot N_{1.5} - K_{2.5}^i U_{\sigma}^{,i} \right ) +
O(\epsilon^4) \,,\\
\Gamma^0_{0i} &=&
-\epsilon U_{\sigma}^{,i}
-\epsilon^2 \left ( \quarter (N_1^{,i} + B_1^{,i} ) 
	- 4U_{\sigma} U_{\sigma}^{,i} \right
) 
-\epsilon^3 \biggl ( \quarter (N_2^{,i} + B_2^{,i} )  
	- N_1 U_{\sigma}^{,i} 
	- U_{\sigma} N_1^{,i} 
\nonumber \\
&&
	+ 4 V_{\sigma}^i \dot U_{\sigma} 
	+ 8 V_{\sigma}^j V_{\sigma}^{i,j}
	+ 16 U_{\sigma}^2 U_{\sigma}^{,i} \biggr )
- \epsilon^{7/2} \left ( \quarter (N_{2.5}^{,i} + B_{2.5}^{,i} ) -
	N_{1.5} U_{\sigma}^{,i} \right ) 
	+ O(\epsilon^4) \,,\\
\Gamma^0_{ij} &=&
\epsilon^2 (4 V_{\sigma}^{(i,j)} + \dot U_{\sigma} \delta^{ij} ) 
+\epsilon^3 \left ( \quarter (\dot N_1 - \dot B_1) \delta^{ij}
	+ \half \dot B_2^{ij} + K_2^{(i,j)}
	- 16 V_{\sigma}^{(i} U_{\sigma}^{,j)} 
	+4 V_{\sigma}^k U_{\sigma}^{,k} \delta^{ij} 
	\right )
\nonumber \\
&&
+ \epsilon^{7/2} \left ( \half \dot B_{2.5}^{ij} + K_{2.5}^{(i,j)} 
	- \half \dot N_{1.5} \delta^{ij} \right )
+ O(\epsilon^4) \,,\\
\Gamma^i_{00} &=&
- U_{\sigma}^{,i} - \epsilon \left ( \quarter (N_1^{,i} + B_1^{,i} )
	+ 4 \dot V_{\sigma}^i - 8 U_{\sigma} U_{\sigma}^{,i} \right )
- \epsilon^2 \biggl (  \quarter (N_2^{,i} + B_2^{,i} ) + \dot K_2^i
	- 2N_1 U_{\sigma}^{,i} 
\nonumber \\
&&
	- 2U_{\sigma} N_1^{,i} 
	- U_{\sigma} B_1^{,i} - B_2^{ij} U_{\sigma}^{,j}
	- 4 V_{\sigma}^i \dot U_{\sigma}  
	-16 U_{\sigma} \dot V_{\sigma}^i + 8 V_{\sigma}^j V_{\sigma}^{j,i} 
	+48 U_{\sigma}^2 U_{\sigma}^{,i} \biggr )
\nonumber \\
&&
- \epsilon^{5/2} \left ( \quarter (N_{2.5}^{,i} + B_{2.5}^{,i} )
	+ \dot K_{2.5}^i - 2 N_{1.5} U_{\sigma}^{,i} 
	- B_{2.5}^{ij} U_{\sigma}^{,j}
	\right ) 
\nonumber \\
&&
- \epsilon^3 \biggl ( \quarter (N_3^{,i} + B_3^{,i} ) + \dot K_3^i
	- 2 N_2 U_{\sigma}^{,i} - U_{\sigma} (2 N_2^{,i} + B_2^{,i})
	- 4 U_{\sigma} \dot K_2^i - K_2^i \dot U_{\sigma}
\nonumber \\
&&
	- \half N_1 N_1^{,i} - \quarter N_1 B_1^{,i} 
	- V_{\sigma}^i (\dot N_1 + \dot B_1 ) - 4 N_1 \dot V_{\sigma}^i
	+ 2 K_2^j V_{\sigma}^{j,i} + 2 V_{\sigma}^j K_2^{j,i}
	+ 4 V_{\sigma}^j \dot B_2^{ij}
\nonumber \\
&&
	+ \quarter B_2^{jk} B_2^{jk,i} 
	- B_3^{ij} U_{\sigma}^{,j} - \quarter B_2^{ij} ( B_1^{,j} + N_1^{,j} )
	+ 12 U_{\sigma}^2 N_1^{,i} + 4 U_{\sigma}^2 B_1^{,i} 
	+ 24 U_{\sigma} N_1 U_{\sigma}^{,i}
\nonumber \\
&&
	+8 U_{\sigma} U_{\sigma}^{,j} B_2^{ij} 
	+64 U_{\sigma}^2 \dot V_{\sigma}^i 
	+32 U_{\sigma} \dot U_{\sigma} V_{\sigma}^i
	-64 U_{\sigma} V_{\sigma}^j V_{\sigma}^{j,i} 
	- 32 V_{\sigma}^j V_{\sigma}^j U_{\sigma}^{,i} 
	-256 U_{\sigma}^3 U_{\sigma}^{,i} \biggr )
\nonumber \\
&&
-  \epsilon^{7/2} \biggl ( \quarter (N_{3.5}^{,i} + B_{3.5}^{,i} )
	+ \dot K_{3.5}^{i} 
	- \quarter N_{1.5} B_1^{,i} - \half  N_{1.5} N_1^{,i}
	- 2 U_{\sigma} N_{2.5}^{,i} - U_{\sigma} B_{2.5}^{,i} 
	- 2 N_{2.5} U_{\sigma}^{,i}
\nonumber \\
&&
	+ 2 V_{\sigma}^j K_{2.5}^{j,i} + 2 K_{2.5}^j V_{\sigma}^{j,i}
	-4 U_{\sigma} \dot K_{2.5}^i - K_{2.5}^i \dot U_{\sigma}
	- 4 V_{\sigma}^i \dot N_{1.5} - 4 N_{1.5} \dot V_{\sigma}^i
	- U_{\sigma}^{,j} B_{3.5}^{ij} 
	+ 4 V_{\sigma}^j \dot B_{2.5}^{ij}
\nonumber \\
&&
	- \quarter B_{2.5}^{ij} ( N_1^{,j} + B_1^{,j} )
	+ \quarter B_{2.5}^{jk} B_2^{jk,i} 
	+24 U_{\sigma} N_{1.5} U_{\sigma}^{,i} 
	+ 8 U_{\sigma} U_{\sigma}^{,j} B_{2.5}^{ij} \biggr )
 + O(\epsilon^4) \,,  \\
\Gamma^i_{0j} &=&
\epsilon ( \dot U_{\sigma} \delta^{ij} - 4 V_{\sigma}^{[i,j]} )
+ \epsilon^2 \biggl ( \quarter (\dot N_1 -\dot B_1) \delta^{ij}
	- K_2^{[i,j]} + \half \dot B_2^{ij}  
	-4 U_{\sigma} \dot U_{\sigma} \delta^{ij} 
\nonumber \\
&&
	- 4 V_{\sigma}^j U_{\sigma}^{,i} + 16 U_{\sigma} V_{\sigma}^{[i,j]} 
	\biggr )
- \epsilon^{5/2} \left ( \half \dot N_{1.5} \delta^{ij} + K_{2.5}^{[i,j]}
	- \half \dot B_{2.5}^{ij} \right )
\nonumber \\
&&
+ \epsilon^3 \biggl ( \quarter (\dot N_2 - \dot B_2 ) \delta^{ij}
	- K_3^{[i,j]} + \half \dot B_3^{ij}
	- (N_1 \dot U_{\sigma} + U_{\sigma} \dot N_1) \delta^{ij}
	- K_2^j U_{\sigma}^{,i} + 4U_{\sigma} K_2^{[i,j]} 
\nonumber \\
&&
	- 16 V_{\sigma}^{(i} \dot V_{\sigma}^{j)} 
	+ 8 V_{\sigma}^{k} \dot V_{\sigma}^{k} \delta^{ij}
	- V_{\sigma}^j (N_1^{,i} + B_1^{,i}) + 4 N_1 V_{\sigma}^{[i,j]}
	+ 4 B_2^{ik} V_{\sigma}^{[k,j]} 
\nonumber \\
&&
	+ 4 V_{\sigma}^{k,[i} B_2^{j]k} 
	- 4 V_{\sigma}^k B_2^{k[i,j]}
	+ 16 U_{\sigma}^2 \dot U_{\sigma} \delta^{ij}
	+ 32 U_{\sigma} U_{\sigma}^{,i} V_{\sigma}^j 
	- 64 U_{\sigma}^2 V_{\sigma}^{[i,j]} \biggr )
\nonumber \\
&&
+ \epsilon^{7/2} \biggl ( \quarter (\dot N_{2.5} - \dot B_{2.5} )\delta^{ij}
	-  K_{3.5}^{[i,j]} + \half \dot B_{3.5}^{ij}
	- (N_{1.5} \dot U_{\sigma} + U_{\sigma} \dot N_{1.5} ) \delta^{ij}
\nonumber \\
&&
	- K_{2.5}^j U_{\sigma}^{,i} + 4 U_{\sigma} K_{2.5}^{[i,j]}
	+ 4 N_{1.5} V_{\sigma}^{[i,j]}
	+ 4 B_{2.5}^{ik} V_{\sigma}^{[k,j]} + 4 V_{\sigma}^{k,[i} B_{2.5}^{j]k} 
	\biggr )
+ O(\epsilon^4) \,, \\
\Gamma^i_{jk} &=&
\epsilon \,{^0}\Gamma^i_{jk}(U_{\sigma})
+ \epsilon^2 \left ( \quarter \,{^0}\Gamma^i_{jk} (N_1 - B_1)
	-4 U_{\sigma} \,{^0}\Gamma^i_{jk}(U_{\sigma}) 
	+ \half (B_2^{ij,k} + B_2^{ik,j} - B_2^{jk,i} ) \right )
\nonumber \\
&&
+ \epsilon^{3}  \biggl ( \quarter \,{^0}\Gamma^i_{jk} (N_2 - B_2)
	+ \half (B_3^{ij,k} + B_3^{ik,j} - B_3^{jk,i} )
	- N_1 \,{^0}\Gamma^i_{jk}(U_{\sigma}) 
	+ B_2^{il} U_{\sigma}^{,l} \delta^{jk} - B_2^{jk} U_{\sigma}^{,i}
\nonumber \\
&&
	- U_{\sigma} \,{^0}\Gamma^i_{jk}(N_1)
	+ 16 U_{\sigma}^2 \,{^0}\Gamma^i_{jk}(U_{\sigma}) 
	+ 4 V_{\sigma}^i \dot U_{\sigma} \delta^{jk} 
	+ 4 \,{^0}\Gamma^i_{jk}(V_{\sigma}^l V_{\sigma}^l)
	- 16 V_{\sigma}^j V_{\sigma}^{[i,k]} 
	- 16 V_{\sigma}^k V_{\sigma}^{[i,j]} \biggr )
\nonumber \\
&&
+ \epsilon^{7/2} \biggl ( \quarter \,{^0}\Gamma^i_{jk} (N_{2.5} - B_{2.5})
	+ \half (B_{3.5}^{ij,k} + B_{3.5}^{ik,j} - B_{3.5}^{jk,i} )
\nonumber \\
&&
	- N_{1.5} \,{^0}\Gamma^i_{jk}(U_{\sigma})
	+ B_{2.5}^{il} U_{\sigma}^{,l} \delta^{jk} 
	- B_{2.5}^{jk} U_{\sigma}^{,i}
	\biggr ) + O(\epsilon^4) \,,
\end{eqnarray}
\label{gammas}
\end{mathletters}
where we define
\begin{equation}
{^0}\Gamma^i_{jk}(f) \equiv
 f^{,k} \delta^{ij} + f^{,j} \delta^{ik} - f^{,i} \delta^{jk} \,.
\end{equation}

\subsection{Conversion to the baryon density $\rho^*$}

We must now convert all potentials from integrals over $\sigma$, $\sigma^i$
and $\sigma^{ij}$ to integrals over the conserved baryon density $\rho^*$,
defined by Eq. (\ref{conservedrho}).  
From Eqs. (\ref{sigmadefinitions}) and (\ref{Trhostar}), we find
\begin{eqnarray}
\sigma &=& \rho^* u^0 (1+v^2)/\sqrt{-g} \,,
\nonumber \\
\sigma^i &=& \rho^* u^0 v^i/\sqrt{-g} \,,
\nonumber \\
\sigma^{ij} &=& \rho^* u^0 v^i v^j /\sqrt{-g} \,,
\label{sigmatorho}
\end{eqnarray}
where $u^0 = (-g_{00} -2 g_{0i}v^i - g_{ij}v^iv^j)^{-1/2}$.  Substituting the
expansions for the metric, Eq. (\ref{metricexpand}), and for the metric
potentials Eq. (\ref{expandNKB}), we obtain, to the
order required for 3.5PN equations of motion, 
\begin{mathletters}
\begin{eqnarray}
\sigma &=& \rho^* \biggl [ 1 + 
	\epsilon \left ({3 \over 2} v^2 - U_\sigma \right )
	+ \epsilon^2 \left ( {7 \over 8} v^4 + \half v^2 U_\sigma 
	- 4 v^j V_\sigma^j  - \quarter N_1
	+ {3 \over 4} B_1 + {5 \over 2} U_\sigma^2 \right )
\nonumber \\
&&
	+ 2 \epsilon^{5/2} N_{1.5}
	+ \epsilon^3 \left ( {11 \over 16} v^6 + {33 \over 8} v^4 U_\sigma
	-10 v^2 V_\sigma^j v^j + {7 \over 4} v^2 U_\sigma^2
	+ {1 \over 8} v^2 N_1 + {9 \over 8} v^2 B_1 + \half B_2^{ij} v^i v^j
	\right .
\nonumber \\
&&
	\left .
	+ 4 U_\sigma V_\sigma^j v^j - K_2^j v^j
	- \quarter N_2 + {3 \over 4} B_2 
	- 4 V_\sigma^i V_\sigma^i
	+ {5 \over 4} U_\sigma N_1
	- {3 \over 4} U_\sigma B_1 
	-{15 \over 2} U_\sigma^3 
\right ) 
\nonumber \\
&&
	+ \epsilon^{7/2} \left ({7 \over 2} v^2 N_{1.5} 
	+ \half B_{2.5}^{ij} v ^i v^j - K_{2.5}^j v^j
	- \quarter N_{2.5}
	+ {3 \over 4} B_{2.5}- U_\sigma N_{1.5}  
	\right )
	+ O(\epsilon^4) \biggr ] \,, \\
\sigma^i &=& \rho^*  v^i \biggl [ 1 + 
	\epsilon \left ( \half v^2 - U_\sigma \right)
	+\epsilon^2 \left ({3 \over 8} v^4 + {3 \over 2} v^2 U_\sigma 
	- 4 V_\sigma^j v^j
	+ {3 \over 4} B_1 - \quarter N_1 + {5 \over 2} U_\sigma^2 \right )
\nonumber \\
&&
	+ 2 \epsilon^{5/2} N_{1.5}
	+ O(\epsilon^3) \biggr ] \,, \\
\sigma^{ij} &=& \rho^* v^i v^j \biggl [ 1 +
	\epsilon \left ( \half v^2 - U_\sigma \right) 
	+ O(\epsilon^2) \biggr ] \,, \\
\sigma^{ii} &=& \rho^* v^2 \biggl [ 1 +
	\epsilon \left ({1 \over 2} v^2 - U_\sigma \right )
	+\epsilon^2 \left ({3 \over 8} v^4 + {3 \over 2} v^2 U_\sigma 
	- 4 V_\sigma^j v^j
	+ {3 \over 4} B_1 - \quarter N_1 + {5 \over 2} U_\sigma^2 \right )
\nonumber \\
&&
	+2 \epsilon^{5/2} N_{1.5}
	+ O(\epsilon^3) \biggr ] \,. 
\end{eqnarray}
\label{sigmatorhoPN}
\end{mathletters}
Substituting these formulae into the definitions for $U_\sigma$ and the
other potentials defined in Paper I, Eqs. (4.10) -- (4.16), and iterating
successively, we convert all such potentials into  new potentials defined
using $\rho^*$, plus PN corrections.  For example, we find that
\begin{eqnarray}
U_\sigma &=&  U + \epsilon \left ({3 \over 2} \Phi_1 - \Phi_2 \right )  
\nonumber \\
&&
	+ \epsilon^2 \left ( {7 \over 8} \Sigma (v^4) + \half \Sigma (v^2 U)
	-4 \Sigma (v^j V^j) + {5 \over 2} \Sigma (\Phi_1) -\Sigma (\Phi_2)
\right .
\nonumber \\
&&
\left .
	+ {3\over 2} \Sigma (U^2) - \half \Sigma (\ddot X ) \right )
	- \epsilon^{5/2} \left ( {4 \over 3} U \stackrel{(3)}{{\cal I}^{jj}}(t)
	\right )
	+ O(\epsilon^3) \,,
\label{Usigmaexpand}
\end{eqnarray}
where henceforth, $U$, $V^j$, $\Phi_1$, $\Phi_2$, $\Sigma$, and so on, are 
defined
in terms of $\rho^*$ (see Appendix \ref{sec:keyformulae}).

\subsection{Final continuum equations of motion}

Combining Eqs. (\ref{gammas}) and (\ref{geodesic}), substituting the
explicit forms of the potentials $N_1$, $K_2^i$, $B_2^{ij}$, etc. from 
Paper I, Eqs. (5.2), (5.4), (5.8), (5.10), (6.2), and (6.4),
and inserting the iterated forms of all potentials, 
we obtain the equation of motion through 3.5PN order.  
\begin{equation}
dv^i /dt = U^{,i} + a_{PN}^i + a_{2PN}^i + a_{2.5PN}^i + a_{3PN}^i 
+ a_{3.5PN}^i \,,
\end{equation}
where 
\begin{mathletters}
\begin{eqnarray}
a_{PN}^i &=&
	v^2 U^{,i} -4 v^i v^j U^{,j} - 3v^i \dot U - 4 U U^{,i} 
	+ 8 v^j V^{[i,j]} 
	+ 4 \dot V^i + \half \ddot X^{,i} + {3 \over 2} \Phi_1^{,i}
	-\Phi_2^{,i} \,, \\
a_{2PN}^i &=&
 4 v^i v^j v^k V^{j,k} +  v^2 v^i \dot U  
+ v^i v^j ( 4 \Phi_2^{,j} - 2 \Phi_1^{,j} - 2 \ddot X^{,j} )
- \half v^2 (2 \Phi_2^{,i} + \Phi_1^{,i} -  \ddot X^{,i} )
\nonumber \\
&&
+ v^j v^k ( 2 \Phi_1^{jk,i} - 4 \Phi_1^{ij,k} + 2 P_2^{jk,i} -4 P_2^{ij,k})
+v^i ( 3 \dot \Phi_2 - \half \dot \Phi_1 - {3 \over 2} \stackrel{(3)}{X}
	+4 V^k U^{,k} )
\nonumber \\
&&
+ v^j ( 8 V_2^{[i,j]} 
	- 16 \Phi_2^{[i,j]} 
	+ 4 \ddot X^{[i,j]} + 32 G_7^{[i,j]}
	- 16 U V^{[i,j]} 
	- 4 \Sigma^{,[i}(v^{j]}v^2) 
	+ 8 V^i U^{,j} 
\nonumber \\
&&
	- 4 \dot \Phi_1^{ij} 
	- 4 \dot P_2^{ij})
	+ {7 \over 8} \Sigma^{,i}(v^4) 
	+ {9 \over 2} \Sigma^{,i}(v^2 U)
	- 4 \Sigma^{,i}(v^j V^j) 
	- {3 \over 2} \Sigma^{,i}(\Phi_1) 
	- 6 U \Phi_1^{,i} 
	-2 \Phi_1 U^{,i} 
\nonumber \\
&&
	- 4 \Phi_1^{ij} U^{,j} 
	+ 8 V^j V^{j,i} 
	+ 4 V^i \dot U 
	+ 2 \dot \Sigma (v^i v^2)
	+ 4 U \Phi_2^{,i} 
	+ 4 \Phi_2 U^{,i}
	+ 8 U^2 U^{,i} 
	- \Sigma^{,i}(\Phi_2) 
\nonumber \\
&&
	+ {3 \over 2} \Sigma^{,i}(U^2) 
	-2 U \ddot X^{,i} 
	- 2 \ddot X U^{,i}
	- 8 U \dot V^i 
	- \half \Sigma^{,i}(\ddot X )
	+ {3 \over 4} \ddot X_1^{,i} 
	- \half \ddot X_2^{,i}
	+ 2 \stackrel{(3)}{X^{i}}
	+ {1 \over 24} \stackrel{(4)}{Y^{,i}} 
\nonumber \\
&&
	+ 4 \dot V_2^i 
	- 8 \dot \Phi_2^i 
	- 6G_1^{,i} 
	- 4 G_2^{,i} 
	+ 8 G_3^{,i} 
	+ 8 G_4^{,i} 
	- 4 G_6^{,i} 
	+ 16 \dot G_7^i 
	- 4 P_2^{ij} U^{,j}
	- 4 H^{,i}
\,, \\
a_{2.5PN}^i &=&  {3 \over 5} x^j ( \stackrel{(5)}{{\cal I}^{ij}} - {1 \over 3}
\delta ^{ij} \stackrel{(5)}{{\cal I}^{kk}} ) + 
2 v^j \stackrel{(4)}{{\cal I}^{ij}}
+ 2 U^{,j} \stackrel{(3)}{{\cal I}^{ij}} 
+ {4 \over 3} U^{,i} \stackrel{(3)}{{\cal I}^{kk}}
- X^{,ijk} \stackrel{(3)}{{\cal I}^{jk}}
\nonumber \\
&&
- {2 \over 15} \stackrel{(5)}{{\cal I}^{ijj}}
+ {2 \over 3} \epsilon^{qij} \stackrel{(4)}{J^{qj}} \,, \\
a_{3.5PN}^i &=&
{1 \over 210} (13r^2x^k \delta^{ij}-4r^2x^i \delta^{jk} - x^ix^jx^k) 
	 \stackrel{(7)}{{\cal I}^{jk}}
\nonumber \\
&&
+{1 \over 30} ( 10r^2v^k \delta^{ij} + 4(v \cdot x)x^k \delta^{ij}
	-r^2v^i \delta^{jk} -4x^ix^jv^k + 3x^jx^kv^i ) 
	\stackrel{(6)}{{\cal I}^{jk}}
\nonumber \\
&&
+{1 \over 15} ( 10(v \cdot x)v^k \delta^{ij} - x^kv^2 \delta^{ij}
	+2(v \cdot x)v^i \delta^{jk} +2x^iv^2 \delta^{jk}
	-5x^iv^jv^k + 4x^jv^iv^k )
	\stackrel{(5)}{{\cal I}^{jk}}
\nonumber \\
&&
+{1 \over 15} (5r^2U^{,j} - 35x^jU + 59X^{,j} ) 
\stackrel{(5)}{{\cal I}^{ij}}
+{1 \over 30} (5r^2U^{,i} + 30x^iU - 34X^{,i} - 6x^jX^{,ij})
	\stackrel{(5)}{{\cal I}^{kk}}
\nonumber \\
&&
+{1 \over 90} (15x^jx^kU^{,i} - 6x^jX^{,ik} - 30x^iX^{,jk}
	-15r^2X^{,ijk} - 4Y^{,ijk} + 5x^lY^{,ijkl} )
 	\stackrel{(5)}{{\cal I}^{jk}}
\nonumber \\
&&
+{1 \over 3}v^i ( v^2 \delta^{jk} - v^jv^k) \stackrel{(4)}{{\cal I}^{jk}}
-{1 \over 3} (2x^j{\dot U} + 22V^j - 14 {\dot X}^{,j} - 12v^kX^{,jk})
	\stackrel{(4)}{{\cal I}^{ij}}
\nonumber \\
&&
+{1 \over 18} (24x^jv^kU^{,i}+12v^ix^jU^{,k} + 12x^jV^{k,i}
	+54v^iX^{,jk} -72v^jX^{,ik} 
\nonumber \\
&&
	+ 12x^j {\dot X}^{,ik}
	+60X^{j,ki} - 72X^{i,jk} - 5{\dot Y}^{,ijk} )
	\stackrel{(4)}{{\cal I}^{jk}}
	-{1 \over 3} (6v^iU-8V^i-2{\dot X}^{,i}) 
	\stackrel{(4)}{{\cal I}^{kk}}
\nonumber \\
&&
+(2v^2U^{,j}- 8UU^{,j} -8v^kV^{k,j}+3\Phi_1^{,j} -2\Phi_2^{,j}
	+{\ddot X}^{,j} ) \stackrel{(3)}{{\cal I}^{ij}}
\nonumber \\
&&
+{1 \over 3} (2v^2U^{,i} - 8v^iv^jU^{,j} - 24UU^{,i} -6v^i{\dot U}
	+8 {\dot V}^i + 16v^jV^{[i,j]} +3\Phi_1^{,i} -6\Phi_2^{,i}
	+{\ddot X}^{,i} ) \stackrel{(3)}{{\cal I}^{kk}}
\nonumber \\
&&
+{1 \over 6} (48v^jV^{k,i} -12v^jv^kU^{,i} - 6v^2X^{,ijk}
	+24v^iv^lX^{,jkl} 
	-24{\dot X}^{i,jk}
	-9X_1^{,ijk} 
	+ 6X_2^{,ijk} 
\nonumber \\
&&
	+18v^i{\dot X}^{,jk} 
	- 48v^lX^{[i,l]jk} 
	+24UX^{,ijk} +24U^{,i}X^{,jk}
	- 18\Phi_1^{jk,i} 
	+6\Sigma^{,i}(X^{,jk}) - {\ddot Y}^{ijk} )
	\stackrel{(3)}{{\cal I}^{jk}}
\nonumber \\
&&
+{1 \over 630} (10x^ix^j \delta^{kl} - 25 x^kx^l \delta^{ij}
	- 9r^2 \delta^{ij}\delta^{kl} )  \stackrel{(7)}{{\cal I}^{jkl}}
\nonumber \\
&&
+{1 \over 45} (4x^{[i}v^{j]} \delta^{kl} - 2(v \cdot x)
	\delta^{ij}\delta^{kl} -10x^kv^l \delta^{ij} )
	\stackrel{(6)}{{\cal I}^{jkl}}
\nonumber \\
&&
-{1 \over 45} (v^2 \delta^{ij}\delta^{kl} + 6v^iv^j \delta^{kl}
	+ 5 v^kv^l \delta^{ij} ) \stackrel{(5)}{{\cal I}^{jkl}}
+{1 \over 9} (4U \delta^{jk} -2x^jU^{,k}+X^{,jk} )
	\stackrel{(5)}{{\cal I}^{ijk}}
\nonumber \\
&&
+{1 \over 45} (4X^{,ij} - 10x^jU^{,i} )\stackrel{(5)}{{\cal I}^{jkk}}
+{1 \over 54} (6x^jX^{,ikl} - Y^{,ijkl} ) \stackrel{(5)}{{\cal I}^{jkl}}
\nonumber \\
&&
+{2 \over 9} ({\dot U}\delta^{ij}-v^iU^{,j}-2v^jU^{,i} -V^{j,i}
	-{\dot X}^{,ij} ) \stackrel{(4)}{{\cal I}^{jkk}}
\nonumber \\
&&
+{1 \over 45} (3r^2 \epsilon^{qik} + 2x^ix^j \epsilon^{qjk}
	+4x^jx^k \epsilon^{qij} ) \stackrel{(6)}{{\cal J}^{qk}}
-{16 \over 45} x^jv^k \epsilon^{qjk} \stackrel{(5)}{{\cal J}^{qi}}
\nonumber \\
&&
+{2 \over 45} (2(v \cdot x) \epsilon^{qik} -2x^iv^j \epsilon^{qjk}
	+ 5x^jv^i \epsilon^{qjk} + 12x^jv^k \epsilon^{qij}
	+4x^kv^j \epsilon^{qij} )\stackrel{(5)}{{\cal J}^{qk}}
\nonumber \\
&&
-{2 \over 9} (13U \epsilon^{qik} +x^j U^{,i} \epsilon^{qjk}
	-2 x^jU^{,k} \epsilon^{qij} 
	-X^{,ij} \epsilon^{qjk}
	-2X^{,jk} \epsilon^{qij}
	-2x^lX^{,ijk} \epsilon^{qlj} ) \stackrel{(4)}{{\cal J}^{qk}}
\nonumber \\
&&
+{2 \over 9} (4v^jv^k \epsilon^{qij} - v^2 \epsilon^{qik} )
	\stackrel{(4)}{{\cal J}^{qk}}
-{4 \over 9} x^j U^{,k} \epsilon^{qjk} \stackrel{(4)}{{\cal J}^{qi}}
-{4 \over 9} {\dot U} \epsilon^{qik}\stackrel{(3)}{{\cal J}^{qk}}
\nonumber \\
&&
+{4 \over 9} (v^iU^{,j}+2v^jU^{,i}+V^{j,i}+{\dot X}^{,ij} )
	\epsilon^{qjk}\stackrel{(3)}{{\cal J}^{qk}}
-{1 \over 840} x^i \stackrel{(7)}{{\cal I}^{jjkk}}
\nonumber \\
&&
+{1 \over 35} x^j \stackrel{(7)}{{\cal I}^{ijkk}}
+{1 \over 40} v^i \stackrel{(6)}{{\cal I}^{jjkk}}
+{1 \over 24} U^{,i} \stackrel{(5)}{{\cal I}^{jjkk}}
-{1 \over 30} x^j \epsilon^{qij} \stackrel{(6)}{{\cal J}^{qkk}}
-{1 \over 15} x^j \epsilon^{qik} \stackrel{(6)}{{\cal J}^{qjk}}
\nonumber \\
&&
+{1 \over 15} v^j (\epsilon^{qjk}\stackrel{(5)}{{\cal J}^{qik}}
	-\epsilon^{qik} \stackrel{(5)}{{\cal J}^{qjk}} 
	-\epsilon^{qij}\stackrel{(5)}{{\cal J}^{qkk}} )
-{1 \over 30} x^i \stackrel{(5)}{{\cal M}^{kkjj}} 
-{1 \over 15} x^j \stackrel{(5)}{{\cal M}^{kkij}} 
\nonumber \\
&&
-{1 \over 6} v^i \stackrel{(4)}{{\cal M}^{kkjj}} 
+{2 \over 3} v^j \stackrel{(4)}{{\cal M}^{ijkk}}
+{1 \over 6}U^{,i}\stackrel{(3)}{{\cal M}^{jjkk}}
+{2 \over 3} U^{,j} \stackrel{(3)}{{\cal M}^{ijkk}}
\nonumber \\
&&
-{1 \over 3} X^{,ijk} \stackrel{(3)}{{\cal M}^{jkll}}
-{23 \over 4200} \stackrel{(7)}{{\cal I}^{ijjkk}}
+{2 \over 75} \epsilon^{qij}\stackrel{(6)}{{\cal J}^{qjkk}}
+{1 \over 30} \stackrel{(5)}{{\cal M}^{kkjji}} \,.
\end{eqnarray}
\label{eomfluid}
\end{mathletters}
Because of their length, we 
shall defer presentation of the 3PN contributions to later
publications when they will actually be needed for calculations.  

\section{Two-body equations of motion to 2PN order}
\label{sec:2body}

\subsection{General treatment of ``spherical pointlike'' masses}

We must now integrate all potentials that appear in the equation of motion,
as well as the equation of motion (\ref{eomfluid}) itself over the bodies in
the binary system.  We treat each body as a non-rotating, spherically symmetric
fluid ball (as seen in its momentary rest frame), whose characteristic
size $s$ is much smaller than the orbital separation.  We shall
discard all terms in the resulting equations that are proportional to
positive powers of $s$: these correspond to multipolar interactions and
their relativistic corrections.  The leading
Newtonian quadrupole 
effect is formally of order $(s/r)^2$ relative to the monopole gravitational
potential $m/r$, but for compact objects such as neutron stars or
black holes, $s \sim m$, so {\it effectively} this is comparable to a
2PN term.  Furthermore, if the quadrupole moment is the result of tidal
interaction with the companion, the size of the induced moment is of order
$(s/r)^3$, so the net effect is $O(s/r)^5$, or roughly 5PN order.  
Such leading multipolar terms can be calculated straightforwardly,
but here we ignore them.  

We also discard all terms that are proportional to negative powers of
$s$: these correspond to self-energy corrections of PN and higher
order.  We shall assume that all such corrections can be merged
uniformly into a suitably renormalized mass for each body, in line
with the Strong Equivalence Principle.  
This should be checked by direct calculation, but
here we ignore such terms.

We retain only terms that are proportional to $s^0$.  For the most
part, these are the expected terms that depend on the two masses, terms
that
one would have obtained from a ``delta-function'' approach that
discarded all divergent self-energy terms.  However, at higher PN
orders, another class of $s^0$ terms is possible, at least in
principle.  These are terms that arise from non-linear combinations of
potentials.  One could imagine one potential being expanded in a
multipolar expansion about the center of mass of one of the bodies
in positive powers of $s$, multiplied by another
potential which
 is a ``self-energy'' potential of that body, dependent upon negative
powers of $s$.  One could then end up with a term that has a piece that is
independent of the scale size $s$ of the body, but that still
depends on its internal density distribution.   We will show that such
terms cannot appear at 1PN order by a simple symmetry argument.  At
2PN order, terms of this kind {\it could} appear in certain non-linear
potentials, but in fact 
vanish identically by a subtler symmetry.  At 3PN order, such $s^0$
terms definitely appear, but whether they survive in the final
equations of motion
is an open question at present.  We will discuss
these matters explicitly at each PN order.   This approach is, in some
sense, a ``quick and dirty'' version of the Hadamard {\it partie
finie} technique, but with the virtue that finite-size or
structure-dependent terms can in
principle be
systematically kept and examined.

Our assumption that the bodies are non-rotating will imply simply that
every element of fluid in the body has the same coordinate velocity,
so that $v^i$ can be pulled outside any integral.  This assumption can
be easily modified in order to deal, for example, with rotating
bodies.

Finally we assume that each body is suitably spherical.  By this we
mean that, in a local inertial frame comoving with
the body and centered at its baryonic center of mass, the baryon density
distribution is static and spherically symmetric in the coordinates of
that frame.  In Appendix \ref{sec:spherical}, we show that the transformation
between our global harmonic coordinates $x^i$ and the spatial coordinates
${\hat x}^i$ of this frame can be written in the form
\begin{equation}
x^i = x_A^i + {\hat x}^j \{\delta^i_j + \epsilon ( A^i_j  + B^i_{jk}
	 {\hat x}^k) + O(\epsilon^2) \} \,,
\label{flattening}
\end{equation}
where the subscript $A$ refers to the Ath body and $x_A^j$ denotes its
baryonic center of mass.  The coefficients
$A^i_j$ include the effect of Lorentz boosts, and the
coefficients $B^i_{jk}$ depend on the 
acceleration of the frame in the field of the companion star.
Then, in terms of the local coordinates ${\hat x}^i$, we assume that the baryon
density is spherically symmetric and static, so that $\rho^*
({\hat t}, \hat {\bf x}) = \rho_S (\hat r)$.  As a consequence, a finite-size
moving body will no longer appear spherical, in part because of the
Lorentz-FitzGerald contraction.  This will result in relativistically
induced multipole moments for the body, albeit of order $\epsilon$
relative to the monopole moment.  Ordinarily, these would result in
terms of positive powers of $s$ in the equations of motion, which 
we ignore (in other words, as
the body's size shrinks to zero, the flattening become irrelevant);
however, as before, in terms with products of  potentials, we must worry
about the effect of self-potentials with negative powers of $s$ offsetting
the positive powers from the flattening. 
We show in the Appendix, however that no such terms arise in the
equations of motion at 2PN order, but that they will contribute in
principle at 3PN
order.

\subsection{Newtonian and PN terms}

We shall evaluate the acceleration consistently for body \#1;  the
corresponding equation for body \#2 can be obtained by interchange.
At the end, we shall find the centre-of-mass and relative equations of
motion.

The Newtonian acceleration is straightforward:
\begin{eqnarray}
{(a^i_1)}_N &=& -(1/m_1) \int_1  \rho^* d^3x \int {\rho^*}^\prime {{(x^i -
{x^i}^\prime)} \over {|{\bf x} -
{\bf x}^\prime |^3}} d^3x^\prime 
\nonumber \\
&=&  
-(1/m_1) \int_1 \int_1  \rho^* {\rho^*}^\prime {{(x^i -
{x^i}^\prime)} \over {|{\bf x} -
{\bf x}^\prime |^3}} d^3x d^3x^\prime  -
 (1/m_1) \int_1  \rho^* d^3x \int_2 {\rho^*}^\prime {{(x^i -
{x^i}^\prime)} \over {|{\bf x} -
{\bf x}^\prime |^3}} d^3x^\prime  \,.
\label{newtonterm} 
\end{eqnarray}
The first term vanishes by symmetry, irrespective of any relativistic
flattening or any other effect (Newton's third law).  
Substituting Eq. (\ref{flattening}) 
for each body 
and expanding the second term in powers of ${\hat x}$, using the
general formula
\begin{equation}
{1 \over {|{\bf x} + {\bf y}|}} = \sum_{q=0}^\infty {y^Q \over q!}
	\nabla^Q \left ( {1 \over r} \right ) \,, 
	\quad |{\bf y}| < |{\bf x}| = r \,,
\label{multipoleexpand}
\end{equation}
we find that all contributions apart from the leading term
are of positive powers in $s$,
including the effects of relativistic flattening, and thus are
dropped, with the result
\begin{equation}
{(a^i_1)}_N  = - m_2 n^i/r^2 \,,
\end{equation}
where we define 
${\bf x} = {\bf x}_1 - {\bf x}_2$, $r = |{\bf x}|$, ${\bf n} = {\bf x}/r$.  

The 1PN terms are similarly straightforward. A term such as $v^2 U^{,i}$ is 
integrated over body 1 by setting $v=v_1$ and writing $U=U_1 + U_2$.  With
$v^2$ pulled outside the integral, the integration is equivalent to that of
the Newtonian term (\ref{newtonterm}), with the result
$v^2 U^{,i} \to -m_2 v_1^2 n^i /r^2$.  Other 1PN terms involving quadratic
powers of velocity ($v^i \dot U$, $v^j V^{[i,j]}$, $\Phi_1^{,i}$ and the
velocity-dependent parts of ${\dot V}^i$ and ${\ddot X}^{,i}$ ) are treated
similarly.  Relativistic flattening plays no role through 3PN order.

In the non-linear 
term $UU^{,i}$, the term involving 
$U_1 U_1^{,i}$ is of order $s^i/s^4$, where $s^i$ represents a vector, like
$(x-x^\prime)^i$ that resides entirely within the body.  
In Appendix \ref{sec:spherical} 
we argue that
relativistic flattening introduces corrections of order $\epsilon (A+Bs^k)$,
$\epsilon^2 ( C+Ds^k + Es^ks^l)$, with the
generic term scaling as $\epsilon^n (1+s)^n$.  The leading term
($s^i/s^4$) vanishes by spherical symmetry.  Any contributions of overall
order $s^{-2}$, $s^{-1}$ or $s^{+n}$ are 
discarded.  The only way to get a term of order
$s^0$ out of $s^i/s^4$ is 
to have a correction term of order $s^3$, which is automatically of
order $\epsilon^3$, which results in a 4PN term.  
In the  two cross terms $U_1 U_2^{,i}$ and $U_2 U_1^{,i}$, 
$U_1$ and $U_1^{,i}$ are of order $1/s$ and $s^i/s^3$ respectively;
expanding $U_2$ about 
the center of mass of body 1 using Eq. (\ref{multipoleexpand}) 
yields only products of vectors $s^Q$,
including the contributions from relativistic flattening.
Thus the only terms in the product that vary overall 
as $s^0$ will have odd numbers of vectors $s^i$, whose
integral over body \#1 vanishes by spherical symmetry.
Only the term from $U_2 U_2^{,i}$ contributes, and relativistic flattening
produces only corrections of positive powers of $s$.  The result is
$UU^{,i} \to -m_2^2 n^i/r^3$. 

In the 
terms ${\dot V}^i$ and ${\ddot X}^{,i}$, the acceleration $dv^i/dt$ 
appears.  Working to 1PN order, we must insert the Newtonian equation 
of motion; but working to 2PN order (or higher), we must insert the 1PN (or
higher) equations of 
motion; the 2PN terms so generated will be discussed in the next subsection.  
For ${\dot V}^i$, the result 
using the Newtonian equation of motion is
\begin{equation}
{\dot V}^i = - \int \int {\rho^{*\prime} \over 
	{|{\bf x} - {\bf x}^\prime|}} {{\rho^{*\prime\prime} 
	(x^\prime - x^{\prime\prime} )^i } \over 
  	{|{\bf x}^\prime - {\bf x}^{\prime\prime}|}} 
	d^3x^\prime d^3x^{\prime\prime}
	+ \int {{\rho^{*\prime} v^{i \prime} v^\prime 
	\cdot (x - x^{\prime} )} \over 
	{|{\bf x} - {\bf x}^\prime|^3}} d^3x^\prime \,.
\end{equation}
The double integral is integrated over body \#1 similarly 
to the term $UU^{,i}$, and the velocity-dependent term is integrated 
similarly to the term $v^2U^{,i}$.  The general result of
these considerations is that, at 1PN order, only terms are kept in which,
in the quantity ${\bf x}-{\bf x}^\prime$, the two vectors are evaluated at
the baryonic  center of mass of the two different bodies, respectively, and
never within the same body.

The resulting N and 1PN equation of motion is 
\begin{eqnarray}
a_{1\, (PN)}^{i} &=& -{m_2 \over r^2} n^i + {m_2 \over r^2} n^i \left [ 
	4{m_2 \over r} + 5{m_1 \over r} - v_1^2 +4v_1 \cdot v_2 
	- 2v_2^2 + {3 \over 2}(\nvb)^2 \right ]
\nonumber 
\\
&& \quad + {m_2 \over r^2} (v_1 - v_2)^i (4\nvb -3\nva) \,, 
\nonumber 
\\
a_{2\, (PN)}^{i} &=& 1 \rightleftharpoons 2 \,.
\label{1PNeom}
\end{eqnarray}
Under the interchange $1 \rightleftharpoons 2$, $n^i \to -n^i$.  

\subsection{2PN terms}

Since we are only working to 2PN, 2.5PN and 3.5PN orders here, 
we may evaluate the 2PN terms
without regard to the effects of relativistic flattening, since the
corrections would be of 3PN or 4PN order or higher.  We only need to expand any
potentials about the centers of mass of the bodies to identify all
possible $s^0$ terms.
The terms explicitly cubic in $v^i$ [the first two terms 
in Eq. (\ref{eomfluid}b)] 
are simplest to evaluate, since the
potentials involved already appeared at 1PN order.
Integrating over body \#1,
discarding terms that are singular in $s$, we obtain
$a^i_1(1\&2) = -(m_2 / r^2)[4(\v1v2)( \nva )-v_1^2(\nvb ) ]v_1^i$.
The next ten terms are explicitly quadratic in velocities; 
the potentials are both
linear and quadratic in the masses, and some involve time derivatives that
require substituting the Newtonian equation of motion.  Pulling the
velocities outside the integrals leaves potentials similar to those 
that have already been
integrated at 1PN order.  The 
potential $P_2^{ij}$ also appears; unlike the potentials encountered so far
($U$, $V^i$, $X$, $\Phi_2$, $\dots$), which depend on the pairwise separation
between points, {\i.e.} on the distance 
$r_{AB} = |{\bf x}_A - {\bf x}_B|$, 
$P_2^{ij}$ and others like it depend on the distances between points taken
as a triplet, namely on the combination 
$\Delta(ABC) \equiv r_{AB} + r_{AC} + r_{BC}$.  We denote $P_2^{ij}$
and similar potentials like $G_1$, $G_2$ etc. as ``triangle''
potentials.  Their evaluation is discussed 
in Appendix \ref{sec:nonlinearpotentials}.
From these terms only the obvious ``point'' mass
terms arise, while all others are proportional to either negative or
positive powers of $s$.
We obtain, for example, $v^2 \Phi_2^{,i} \to -m_1m_2
v_1^2 n^i/r^3$, $v^jv^k \Phi_1^{ij,k} \to -m_2 v_2^i
(\v1v2)(\nva)/r^2$ and 
$v^jv^k P_2^{ij,k} \to m_1m_2 [n^i(8(\nva)^2-3v_1^2) -5v_1^i(\nva)]/4r^3
-m_2^2 [n^i(4(\nva)^2-v_1^2) -3v_1^i(\nva)]/4r^3 $. 

The next 13 terms in Eq. (\ref{eomfluid}b), which are 
explicitly 
linear in velocities, are quite similar, except that
they involve either an additional time derivative of potentials, or
vector-like potentials ($V_2^i$, $\Phi_2^i$, $G_7^i$).  Triangle
potentials appear in several places ($G_7^i$, ${\dot P}_2^{ij}$).  As before,
expansion about  the centers of mass yields the normal point mass
terms and terms of only positive and negative powers of $s$.

Of the remaining 33 terms, many involve several implicit powers of
velocity coupled to potential-type expressions;  integration of these
terms over
body \#1 is handled as before.  However, several terms do not involve
velocity at all, and thus are cubically non-linear in masses.
Examples include the terms $U^2U^{,i}$, $\Sigma^{,i}(\Phi_2)$, and the
portions of the terms ${\ddot X}U^{,i}$, ${\ddot X}_1^{,i}$,
$Y^{,0000i}$, where time derivatives have generated accelerations,
thence the Newtonian potential.  In these cases, the high-degree of
nonlinearity presents at least the {\it possibility} of new contributions at
order $s^0$.  The term $U^2U^{,i}$ illustrates the issue.  Writing
$U=U_1+U_2$, we have $(U_1^2 + 2U_1U_2 +U_2^2)(U_1^{,i}+U_2^{,i})$, to be
integrated over body \#1.
Consider the term $U_1^2 U_2^{,i}$.  Expanding $U_2$ about the center
of mass of body \#1 using Eq. (\ref{multipoleexpand}) 
and integrating over body \#1, we obtain
\begin{equation}
{1 \over m_1} \int_1 \rho^* d^3 {\bar x} \int_1 {{\rho^{*\prime} 
	d^3 {\bar x}^\prime}
	\over {|{\bf {\bar x}}-{\bf {\bar x}}^\prime|}}
	\int_1  {{\rho^{*\prime\prime} d^3{\bar x}^{\prime\prime}}
        \over {|{\bf {\bar x}}-{\bf {\bar x}}^{\prime\prime}|}}
	\sum_q {{{\bar x}^Q} \over q!} \nabla^Q \nabla^i 
	\left ({1 \over r} \right )
	\,,
\end{equation}
where the ``barred'' coordinates are all defined relative to the center of
mass of body \#1, and $r=|{\bf x}_1 - {\bf x}_2|$.  
Only the term  with $q=2$ can produce a contribution of overall
order $s^0$.  Using the spherical symmetry of body \#1, the result is 
\begin{equation}
{1 \over 6} {1 \over m_1} \left (\int_1 \rho^* {\bar r}^2 U_1^2 
	d^3{\bar x} \right ) \delta^{jk} \nabla^{ijk}
\left ({1 \over r} \right ) 
\propto \nabla^i \nabla^2 \left ({1 \over r} \right ) = 0 \,.
\label{sampleterm}
\end{equation}
Note that the integral in Eq. (\ref{sampleterm}) scales as $s^0$ for a
fixed $m_1$, yet
depends on the internal structure of the body.  Nevertheless,
the term vanishes via a combination of the spherial symmetry of the body,
and the fact that $\nabla^2 (1/r) = 0$.  Other possible $s^0$ terms also
vanish by symmetry, with the final result that $U^2U^{,i} \to -m_2^3n^i/r^4$.
Similarly, for example $\Sigma^{,i}(\Phi_2) \to -m_1m_2^2n^i/r^4$.

Additional cubically nonlinear terms are $P_2^{ij}U^{,j}$, 
and the acceleration-generated terms in $G_2^{,i}$, $G_5^{,i}$
and ${\dot
G}_7{^i}$.  Since these involve the triangle potentials, they will be
discussed in Appendix \ref{sec:nonlinearpotentials};  
they generate no $s^0$ terms,
again because of symmetry combined with $\nabla^2 (1/r) = 0$.
The term $H^{,i} = \nabla^{,i} P(U^{,jk}P_2^{jk})$ 
involves a still more complicated ``quadrangle''
potential, which is a function of four points.  In Appendix
\ref{sec:nonlinearpotentials} we show that it likewise generates
no $s^0$ terms, with the result 
\begin{equation}
(1/m_1) \int_1 \rho^* H^{,i} d^3x = {{m_2 n^i} \over r^4} 
\left ( 2m_1m_2 + {m_2^2 \over 4} \right ) \,.
\label{Hfinal}
\end{equation}

Working to 2PN order, we must also include terms generated by substituting
the 1PN equations of motion into accelerations generated by time derivatives
acting on velocities in 1PN potentials, specifically the terms $4{\dot V}^i
+ {\ddot X}^{,i}/2$.  This leads to the integral
\begin{equation}
{1 \over m_1} \int_1 \rho^* d^3x \int 
	{{\rho^{*\prime} d^3x^\prime} \over {|{\bf x}-{\bf x}^\prime |}}
	\left ( {7 \over 2} \delta^{ij} + {1 \over 2} {\tilde n}^i{\tilde
	n}^j \right ) a_{PN}^j(t,{\bf x}^\prime) \,,
\end{equation}
where ${\tilde n}^i = (x-x^\prime)^i/|{\bf x}-{\bf x}^\prime |$, and we
substitute Eq. (\ref{eomfluid}a) for $a_{PN}^j(t,{\bf x}^\prime)$.
Evaluation of these terms follows the same method already outlined for
normal, non-triangle, 2PN potentials.

The resulting 2PN contributions to the one-body equation of motion are
\begin{eqnarray}
a_{1\,(2PN)}^i &=& \frac{m_2}{r^2}n^i\bigg[
\frac{m_2}{r}\bigg(
4v_2^2
-8v_1\cdot v_2
+2(v_1\cdot n)^2
-4(v_1\cdot n)(v_2\cdot n)
-6(v_2\cdot n)^2
\bigg)
\nonumber\\
&&
+\frac{m_1}{r}\bigg(
\frac{5}{4}v_2^2
-\frac{5}{2}v_1\cdot v_2
-\frac{15}{4}v_1^2
+\frac{39}{2}(v_1\cdot n)^2
-39(v_1\cdot n)(v_2\cdot n)
+\frac{17}{2}(v_2\cdot n)^2
\bigg)
\nonumber \\
&&
-\frac{57}{4}\frac{m_1^2}{r^2}
-\frac{69}{2}\frac{m_1m_2}{r^2}
-9\frac{m_2^2}{r^2}
-2v_2^4
+4v_2^2(v_1\cdot v_2)
-2(v_1\cdot v_2)^2
\nonumber\\
&&
+\frac{3}{2}v_1^2(v_2\cdot n)^2
-6(v_1\cdot v_2)(v_2\cdot n)^2
+\frac{9}{2}v_2^2(v_2\cdot n)^2
-\frac{15}{8}(v_2\cdot n)^4
\bigg]
\nonumber \\
&&
+\frac{m_2}{r^2}(v_1^i - v_2^i)\bigg[
\frac{m_1}{4r}\bigg( {55}(v_2\cdot n)
-{63}(v_1\cdot n)\bigg)
-\frac{2m_2}{r}\bigg((v_1\cdot n)
+(v_2\cdot n)\bigg)
\nonumber \\
&&
+v_1^2(v_2\cdot n)
+4v_2^2(v_1\cdot n)
-5v_2^2(v_2\cdot n)
-6(v_1\cdot n)(v_2\cdot n)^2
\nonumber \\
&&
-4(v_1\cdot n)(v_1\cdot v_2)
+4(v_2\cdot n)(v_1\cdot v_2)
+\frac{9}{2}(v_2\cdot n)^3
\bigg] \,,
\nonumber \\
a_{2\, (2PN)}^{i} &=& 1 \rightleftharpoons 2 \,.
\label{2PNeom}
\end{eqnarray}
These agree
completely with other results \cite{DD81,kopeikin85,GK86,bfp98,futamase01}.

\section{Radiation Reaction to 2.5PN and 3.5PN order}
\label{sec:2bodyreaction}

Since the multipole moments are strictly functions of time,
integrating the 2.5PN and 3.5PN terms 
in Eq. (\ref{eomfluid}) over body \#1 is
straightforward.  The 2.5PN terms are either trivial, involving $x^i$
or $v^i$, or are similar to integrating the Newtonian term.  
The result is:
\begin{eqnarray}
a_{1\, (2.5PN)}^{i} &=&  {3 \over 5} x_1^j ( \stackrel{(5)}{{\cal I}^{ij}} 
	- {1 \over 3} \delta ^{ij} \stackrel{(5)}{{\cal I}^{kk}} ) 
	+ 2 v_1^j \stackrel{(4)}{{\cal I}^{ij}}
	- {1 \over 3} {m_2 \over r^2}n^i \stackrel{(3)}{{\cal I}^{kk}}
	- 3{m_2 \over r^2}n^in^jn^k \stackrel{(3)}{{\cal I}^{jk}}
\nonumber \\
&&
- {2 \over 15} \stackrel{(5)}{{\cal I}^{ijj}}
+ {2 \over 3} \epsilon^{qij} \stackrel{(4)}{J^{qj}} \,, 
\nonumber \\
a_{2\, (2.5PN)}^{i} &=& 1 \rightleftharpoons 2 \,.
\label{2.5PNterms}
\end{eqnarray}
Likewise,
the 3.5PN terms are either trivial, or involve integrating Newtonian
or 1PN-like potentials.  
To keep the expression for the 3.5PN terms simple, we assume that, to lowest
order, $m_1{\bf x}_1 + m_2{\bf x}_2 =0$ (see Sec. 
\ref{sec:relativecoords} for discussion), so
that we can write, in the 3.5PN term only, $x_1^i = (m_2/m) x^i$ 
and $x_2^i = -(m_1/m) x^i$. Defining
$\alpha_2 \equiv m_2/m$, we obtain
\begin{eqnarray}
a_{1\, (3.5PN)}^{i} &=&
{1 \over 210} \alpha_2^3 r^3 (13n^k \delta^{ij}-4n^i \delta^{jk} 
	- n^in^jn^k) 
	 \stackrel{(7)}{{\cal I}^{jk}}
\nonumber \\
&&
+{1 \over 30} \alpha_2^3 r^2 ( 10v^k \delta^{ij} + 4{\dot r} n^k \delta^{ij}
	-v^i \delta^{jk} -4n^in^jv^k + 3n^jn^kv^i ) 
	\stackrel{(6)}{{\cal I}^{jk}}
\nonumber \\
&&
+{1 \over 15} \alpha_2^3 r ( 10{\dot r}v^k \delta^{ij} - n^kv^2 \delta^{ij}
	+2{\dot r} v^i \delta^{jk} +2n^iv^2 \delta^{jk}
	-5n^iv^jv^k + 4n^jv^iv^k )
	\stackrel{(5)}{{\cal I}^{jk}}
\nonumber \\
&&
+{1 \over 15} m_2 \left \{ (55-36\alpha_2)n^k \delta^{ij} 
	- (19-10\alpha_2)n^i \delta^{jk} + (2+6\alpha_2-10\alpha_2^2)n^in^jn^k \right
	\}
	\stackrel{(5)}{{\cal I}^{jk}}
\nonumber \\
&&
+{1 \over 3} \alpha_2^3 v^i ( v^2 \delta^{jk} - v^jv^k) \stackrel{(4)}{{\cal I}^{jk}}
\nonumber \\
&&
+{1 \over 6} {m_2 \over r} \left \{ 42(1-\alpha_2)v^k \delta^{ij}
	-6(3+\alpha_2){\dot r}n^k \delta^{ij} 
	+(7-\alpha_2)v^i \delta^{jk} + (1-\alpha_2){\dot r}n^i \delta^{jk}
	\right .
\nonumber \\
&&
\left . -(19+3\alpha_2)v^in^jn^k +2(15-3\alpha_2-4\alpha_2^2)v^jn^in^k
	-3(5-9\alpha_2+4\alpha_2^2){\dot r}n^in^jn^k  \right \}
	\stackrel{(4)}{{\cal I}^{jk}}
\nonumber \\
&&
+{1 \over 2} {m_2 \over r^2} \left \{ 16(1+\alpha_2){m \over r}n^in^jn^k
	+(2-\alpha_2^2)n^i (4v^jv^k-6v^2n^jn^k)
\right .
\nonumber \\
&&
\left .
	+(3+\alpha_2) n^k (6{\dot r}v^in^j -4v^iv^j)
	+3(1-\alpha_2)^2 n^in^k (5{\dot r}^2n^j -4{\dot r}v^j )
	\right \}
	\stackrel{(3)}{{\cal I}^{jk}}
\nonumber \\
&&
-{1 \over 6} {m_2 \over r^2} \left \{
	8(1-\alpha_2){m \over r}n^i - 2(2-\alpha_2^2)v^2n^i 	
	+2(3+\alpha_2){\dot r}v^i + 3(1-\alpha_2)^2{\dot r}^2n^i 
	\right \}
	\stackrel{(3)}{{\cal I}^{kk}}
\nonumber \\
&&
+{1 \over 630} \alpha_2^2 r^2 (10n^in^j \delta^{kl} - 25 n^kn^l \delta^{ij}
	- 9 \delta^{ij}\delta^{kl} )  \stackrel{(7)}{{\cal I}^{jkl}}
\nonumber \\
&&
+{1 \over 45} \alpha_2^2 r (4n^{[i}v^{j]} \delta^{kl} - 2{\dot r}
	\delta^{ij}\delta^{kl} -10n^kv^l \delta^{ij} )
	\stackrel{(6)}{{\cal I}^{jkl}}
\nonumber \\
&&
-{1 \over 45} \alpha_2^2 (v^2 \delta^{ij}\delta^{kl} + 6v^iv^j \delta^{kl}
	+ 5 v^kv^l \delta^{ij} ) \stackrel{(5)}{{\cal I}^{jkl}}
\nonumber \\
&&
+{1 \over 90} {m_2 \over r} \left \{
	(43 \delta^{kl}+5n^kn^l) \delta^{ij} 
	+(7+10\alpha_2)n^in^j \delta^{kl}
	-15(1-2\alpha_2)n^in^jn^kn^l \right \}
	\stackrel{(5)}{{\cal I}^{jkl}}
\nonumber \\
&&
+{2 \over 9} {m_2 \over r^2} \left \{ v^in^j + 2\alpha_2v^jn^i
	-3(1-\alpha_2){\dot r}n^in^j \right \}
	\stackrel{(4)}{{\cal I}^{jkk}}
\nonumber \\
&&
+{1 \over 45} \alpha_2^2 r^2 (3 \epsilon^{qik} + 2n^in^j \epsilon^{qjk}
	+4n^jn^k \epsilon^{qij} ) \stackrel{(6)}{{\cal J}^{qk}}
-{16 \over 45} \alpha_2^2 r n^jv^k \epsilon^{qjk} \stackrel{(5)}{{\cal J}^{qi}}
\nonumber \\
&&
+{2 \over 45} \alpha_2^2 r (2{\dot r} \epsilon^{qik} -2n^iv^j \epsilon^{qjk}
	+ 5n^jv^i \epsilon^{qjk} + 12n^jv^k \epsilon^{qij}
	+4n^kv^j \epsilon^{qij} )\stackrel{(5)}{{\cal J}^{qk}}
\nonumber \\
&&
+{2 \over 9} \alpha_2^2 (4v^jv^k \epsilon^{qij} - v^2 \epsilon^{qik} )
	\stackrel{(4)}{{\cal J}^{qk}}
-{2 \over 9} {m_2 \over r} \left \{ 
	(1+\alpha_2)n^in^j \epsilon^{qjk} + 10\epsilon^{qik}
	+2n^jn^k \epsilon^{qij} \right \}
	\stackrel{(4)}{{\cal J}^{qk}}
\nonumber \\
&&
-{4 \over 9} {m_2 \over r^2} \left \{ v^in^j + 2\alpha_2v^jn^i 
	-3(1-\alpha_2){\dot r}n^in^j \right \}
	\epsilon^{qjk}\stackrel{(3)}{{\cal J}^{qk}}
\nonumber \\
&&
-{1 \over 840} \alpha_2 r n^i \stackrel{(7)}{{\cal I}^{jjkk}}
+{1 \over 35} \alpha_2 r n^j \stackrel{(7)}{{\cal I}^{ijkk}}
+{1 \over 40} \alpha_2v^i \stackrel{(6)}{{\cal I}^{jjkk}}
-{1 \over 24} {m_2 \over r^2} n^i \stackrel{(5)}{{\cal I}^{jjkk}}
\nonumber \\
&&
-{1 \over 30} \alpha_2 r n^j \epsilon^{qij} \stackrel{(6)}{{\cal J}^{qkk}}
-{1 \over 15} \alpha_2 r n^j \epsilon^{qik} \stackrel{(6)}{{\cal J}^{qjk}}
-{1 \over 15} \alpha_2 v^j (2\epsilon^{q[i|k}\stackrel{(5)}{{\cal J}^{q|j]k}}
	+\epsilon^{qij}\stackrel{(5)}{{\cal J}^{qkk}})
\nonumber \\
&&
-{1 \over 30} \alpha_2 r n^i \stackrel{(5)}{{\cal M}^{kkjj}} 
-{1 \over 15} \alpha_2 r n^j \stackrel{(5)}{{\cal M}^{kkij}} 
-{1 \over 6} \alpha_2 v^i \stackrel{(4)}{{\cal M}^{kkjj}} 
+{2 \over 3} \alpha_2 v^j \stackrel{(4)}{{\cal M}^{ijkk}}
\nonumber \\
&&
+{1 \over 6} {m_2 \over r^2}n^i  ( \stackrel{(3)}{{\cal M}^{jjkk}}
	-6n^jn^k \stackrel{(3)}{{\cal M}^{jkll}} )
-{23 \over 4200} \stackrel{(7)}{{\cal I}^{ijjkk}}
+{2 \over 75} \epsilon^{qij}\stackrel{(6)}{{\cal J}^{qjkk}}
+{1 \over 30} \stackrel{(5)}{{\cal M}^{kkjji}} \,,
\nonumber \\
a_{2\, (3.5PN)}^{i} &=& 1 \rightleftharpoons 2 \,.
\label{3.5PNterms}
\end{eqnarray}

\section{Relative equations of motion}
\label{sec:relative}

\subsection{System center of mass and  the transformation to relative
coordinates}
\label{sec:relativecoords}

It is useful to note that the 1PN equations (\ref{1PNeom}), 
including the 2.5PN terms (\ref{2.5PNterms}),
admit a first integral that 
corresponds to uniform motion of a ``center of mass'' quantity, namely
\begin{equation}
m_1 v_1^i (1+ {1 \over 2} v_1^2) + m_2 v_2^i (1+ {1 \over 2} v_2^2)
 - {1 \over 2} {m_1m_2 \over r} (v_1^j +v_2^j)(\delta^{ij} +
 n^in^j ) - m {\dot {\cal V}}^i = C^i \,.
\label{cmintegral}
\end{equation}
where 
\begin{equation}
{\cal V}^i \equiv - \left ( {2 \over 15}  \stackrel{(3)}{{\cal I}^{ijj}}
        - {2 \over 3} \epsilon^{qik} \stackrel{(2)}{{\cal J}^{qk}}
        \right ) \,,
\label{calV}
\end{equation}
$C^i$ is a constant, and where  we have assumed that, to Newtonian order,
$m_1x_1^i +m_2x_2^i = 0$.  The 2PN corrections to this first integral
will not be needed here.  Equation (\ref{cmintegral}) can also be obtained by
calculating the 
system dipole moment ${\cal I}^i$ to the corresponding order
(see Appendix \ref{sec:dipole}).
Choosing the coordinates 
so that $C^i =0$, 
we obtain the transformation from individual to relative coordinates and 
velocities, 
to 1PN and 2.5PN order,
\begin{eqnarray}
x_1^i &=& {m_2 \over m} x^i + {1 \over 2} \eta {{\delta m} \over m}
(v^2 - {m \over r}) x^i + {\cal V}^i \,, 
\nonumber \\
x_2^i &=& -{m_1 \over m} x^i + {1 \over 2} \eta {{\delta m} \over m}
(v^2 - {m \over r}) x^i + {\cal V}^i \,, 
\nonumber \\
v_1^i &=& {m_2 \over m} v^i + {1 \over 2} \eta {{\delta m} \over m}
\left [ (v^2 - {m \over r}) v^i - {m \over r^2} {\dot r x^i} \right ]
+ {\dot {\cal V}}^i \,,
\nonumber \\
v_2^i &=& -{m_1 \over m} v^i + {1 \over 2} \eta {{\delta m} \over m}
\left [ (v^2 - {m \over r}) v^i - {m \over r^2} {\dot r x^i} \right ]
+ {\dot {\cal V}}^i \,,
\label{transform}
\end{eqnarray}
where $\delta m \equiv m_1 -m_2$. 
These transformations do not affect the Newtonian term, of course.
However, the 1PN and 2.5PN corrections in Eqs. (\ref{transform}) 
will generate 2PN and 3.5PN terms, respectively, 
when we transform the 1PN terms in
the equation of motion to relative coordinates.  The multipole moments
that appear in the 2.5PN terms in the equation of motion (\ref{2.5PNterms}) 
must
also be converted to relative coordinates, keeping any PN corrections
generated by Eqs. (\ref{transform}); this is treated in Appendix
\ref{sec:quadrupole}.  In addition, in the 2.5PN terms,
multiple time derivatives of the multipole moments will generate
accelerations, for which the 1PN relative equations of motion must be
substituted; in the explicitly 3.5PN terms, the Newtonian equation of
motion suffices.

Calculating $a_1^i - a_2^i$ using Eqs. (\ref{1PNeom}),
(\ref{2PNeom}), (\ref{2.5PNterms}) and (\ref{3.5PNterms}),
substituting Eqs. (\ref{transform}) and the time-derivatives of the
multipole moments (\ref{relativemoments}), we obtain 
the final relative equation of motion through 2.5PN order, plus 3.5PN
terms as given in Eqs. (\ref{eomfinal}) and (\ref{eomfinalcoeffs}).  

\subsection{3.5PN radiation reaction and energy-angular-momentum balance}

A useful check of our radiation-reaction terms at 2.5PN and 3.5PN
order is to verify that the resulting energy and angular momentum loss
in the orbital motion is identical to previously derived energy and
angular momentum flux expressions, accurate to 1PN order beyond the
quadrupole approximation.  In fact, Iyer and Will \cite{iyerwill,iyerwill2}
approached this from the opposite direction, beginning with the 1PN
accurate flux expressions \cite{wagwill,bds,junker}, and deriving the most
general form of a two-body relative equation of motion at 2.5PN and
3.5PN order required by energy and angular momentum balance.  Writing
the radiation-reaction terms in the equation of motion (\ref{eomfinal}) in
the general form
\begin{eqnarray}
A_{2.5PN} &=& a_1 v^2 + a_2 m/r + a_3 {\dot r}^2 \,,
\nonumber \\
B_{2.5PN} &=& b_1 v^2 + b_2 m/r + b_3 {\dot r}^2 \,,
\nonumber \\
A_{3.5PN} &=& c_1 v^4 + c_2 v^2m/r + c_3 v^2{\dot r}^2 
	+ c_4{\dot r}^2m/r +c_5 {\dot r}^4 +c_6 (m/r)^2 \,,
\nonumber \\
B_{3.5PN} &=& d_1 v^4 + d_2 v^2m/r + d_3 v^2{\dot r}^2 
        + d_4{\dot r}^2m/r +d_5 {\dot r}^4 +d_6 (m/r)^2 \,,
\label{ABcoeffs}
\end{eqnarray}
they showed that energy and angular momentum balance would hold if and
only if the coefficients $a_i$, $b_i$, $c_i$ and $d_i$ satisfied the
following equations:
\begin{eqnarray}
a_1 &=& 3+3\beta \,, \quad
a_2=23/3 +2\alpha-3\beta\,, \quad
a_3=-5\beta\,, 
\nonumber \\
b_1 &=& 2+\alpha \,, \quad
b_2=2-\alpha\,, \quad
b_3=-3-3\alpha\,,
\label{absolution}
\end{eqnarray}
for the 2.5PN coefficients, and 
\begin{eqnarray}
c_1=&&{1 \over 28} (117+132\eta)-{3 \over 2} \beta (1-3\eta)+
3\delta_2-3\delta_6\,,
\nonumber \\
c_2=&&-{1 \over 42} (297-310\eta)-3\alpha(1-4\eta)-{3 \over 2}
\beta(7+13\eta)
-2\delta_1-3\delta_2+3\delta_5+3\delta_6\,,
\nonumber \\
c_3=&&{5 \over 28}(19-72\eta)+{5 \over 2}
\beta(1-3\eta)-5\delta_2+5\delta_4+5\delta_6\,,
\nonumber \\
c_4=&&-{1 \over 28}(687-368\eta)-6\alpha\eta+{1
\over2}\beta(54+17\eta)-2\delta_2-5\delta_4-6\delta_5\,,
\nonumber \\
c_5=&&-7\delta_4\,,
\nonumber \\
c_6=&&-{1 \over 21}(1533+498\eta)-\alpha(14+9\eta)+3\beta(7+4\eta)
-2\delta_3-3\delta_5\,,
\nonumber \\
d_1=&&-{3 \over 2} (2+\alpha) (1-3\eta)-\delta_1\,,
\nonumber \\
d_2=&&-{1 \over 84}(139+768\eta)-{1 \over 2}\alpha(5+17\eta)
+\delta_1-\delta_3\,,
\nonumber \\
d_3=&&{1 \over 28}(369-624\eta)+{3 \over 2}(3\alpha+2\beta)(1-3\eta)
+3\delta_1-3\delta_6\,,
\nonumber \\
d_4=&&{1 \over 42}(295-335\eta)+{1
\over2}\alpha(38-11\eta)-3\beta(1-3\eta)
+2\delta_1+4\delta_3+3\delta_6\,,
\nonumber \\
d_5=&&{5 \over 28}(19-72\eta)-5\beta(1-3\eta)+5\delta_6\,,
\nonumber \\
d_6=&&-{1 \over 21}(634-66\eta)+\alpha(7+3\eta)+\delta_3\,,
\label{cdsolution}
\end{eqnarray}
for the 3.5PN coefficients.  The two degrees of freedom ($\alpha$,
$\beta$) at 2.5PN order and the six ($\delta_i$) at 3.5PN order
correspond to gauge or coordinate freedom, and have no physical
consequences.  For example, at 2.5PN order, the values $\alpha = -1$,
$\beta=0$ correspond to the gauge used by Damour and Deruelle \cite{DD81},
while the values $\alpha = 4$, $\beta = 5$  correspond to the
so-called Burke-Thorne gauge (see for example \S 36.11 of \cite{MTW}), 
also used by Blanchet \cite{luc93}.  

It is then a non-trivial check of our result to verify that the 18
coefficients in our 2.5PN and 3.5PN terms, Eqs.
(\ref{eomfinalcoeffs2.5PN}) and (\ref{eomfinalcoeffs3.5PN})  yield a unique,
self-consistent solution for the 8 gauge coefficients.  The result is
\begin{eqnarray}
\alpha &=& -1 \,, \quad \beta=0 \,,
\nonumber \\
\delta_1 &=& {271 \over 28} +6\eta \,, 
\quad \delta_2=-{77 \over 4} - {3 \over 2}\eta \,,
\quad \delta_3={79 \over 14}-{92 \over 7} \eta \,, 
\nonumber \\
\delta_4 &=& 10 \,, 
\quad \delta_5={5 \over 42}+ {242 \over 21}\eta \,, 
\quad \delta_6=-{439 \over 28} +{18 \over 7} \eta \,.
\label{deltacheck}
\end{eqnarray}

\section{Concluding remarks}
\label{sec:conclusions}

We have successfully used DIRE to derive equations of motion for compact
binary systems through 2.5PN and to 3.5PN order, with results consistent
with other methods.  
Instead of using formal delta-function, matching, or regularization
techniques to treat the bodies, 
we modeled them as fluid balls, considered to be
suitably spherical and non-rotating, and small compared to their separation,
and carried out explicit integrations over them.  We then used a technique
whereby we could identify those contributions to the equation of motion that
are independent of the scale size of the bodies (for given masses).
This method can be extended straightforwardly to the complicated 3PN
contributions, to spinning bodies, and to bodies with tidal interactions.
We can also consider the effects at higher PN order of internal self-gravity
(contributions of order $s^{-n})$.  These are the subjects of ongoing
research.

\acknowledgments

This work is supported in part by the National Science Foundation under
grant numbers PHY 96-00049 and PHY 00-96522.   We especially acknowledge
the contribution of Ken Hsieh, who carefully verified a number 
of the calculations
in this paper.

\appendix

\section{Key formulae used in the equations of motion}
\label{sec:keyformulae}

Here we summarize some of the key formulae from Paper I
\cite{patiwill} that will be
needed here.  The potentials that appear in the equations of motion
are all Poisson-like potentials and their generalizations, namely a
superpotential and a superduperpotential:
\begin{eqnarray}
P(f) &\equiv& {1 \over {4\pi}} \int_{\cal M} {{f(t,{\bf x}^\prime)}
\over {|{\bf x}-{\bf x}^\prime | }} d^3x^\prime \,, \quad \nabla^2
P(f) = -f \,, \nonumber \\
S(f)&\equiv& {1 \over {4\pi}} \int_{\cal M}f(t,{\bf x}^\prime)|{\bf
x}-{\bf x}^\prime | d^3x^\prime \,, \quad \nabla^2 S(f) = 2P(f) \,,
\nonumber \\
SD(f)&\equiv&  {1 \over {4\pi}} \int_{\cal M}f(t,{\bf x}^\prime)|{\bf
x}-{\bf x}^\prime |^3 d^3x^\prime \,, \quad \nabla^2 SD(f) = 12S(f)
\,.
\label{definepoisson}
\end{eqnarray}
Note that, in evaluating Poisson potentials
and superpotentials of sources that do not have compact support, our
rule is to evaluate them on the finite, constant time hypersurface
$\cal M$, and to discard all terms that depend on the radius of the
near-zone, $\cal R$.
Unlike Paper I, we now define all potentials in terms of the conserved
baryon density $\rho^*$:
\begin{eqnarray}
\Sigma (f) &\equiv& \int_{\cal M} {{\rho^*(t,{\bf x}^\prime)f(t,{\bf
x}^\prime)}
\over {|{\bf x}-{\bf x}^\prime | }} d^3x^\prime = P(4\pi\rho^* f) \,,
\nonumber \\
X(f)  &\equiv& \int_{\cal M} {\rho^*(t,{\bf x}^\prime)f(t,{\bf
x}^\prime)}
{|{\bf x}-{\bf x}^\prime | } d^3x^\prime = S(4\pi\rho^* f) \,,
\nonumber \\
Y(f) &\equiv& \int_{\cal M} {\rho^*(t,{\bf x}^\prime)f(t,{\bf
x}^\prime)}
{|{\bf x}-{\bf x}^\prime |^3 } d^3x^\prime = SD(4\pi\rho^* f) \,.
\label{definesuper}
\end{eqnarray}

The specific potentials used in the 2PN, 2.5PN and 3.5PN equations of
motion are then given by
\begin{eqnarray}
U &\equiv& \Sigma(1) \,,  \qquad V^i \equiv \Sigma(v^i) \,, \qquad  
\Phi_1^{ij} \equiv \Sigma(v^iv^j)
\,,
\nonumber \\
\Phi_1 &\equiv& \Sigma(v^2) \,, \qquad
\Phi_2 \equiv \Sigma(U) \,, \qquad  X \equiv X(1) \,,
\nonumber \\
V_2^i &\equiv& \Sigma(v^iU) \,, \qquad  \Phi_2^i \equiv \Sigma(V^i) \,,
\qquad Y \equiv Y(1) \,,
\nonumber \\
X^i &\equiv& X(v^i) \,, \qquad  
X_1 \equiv  X(v^2) \,, \qquad  X_2 \equiv X(U) \,,
\nonumber \\
P_2^{ij} &\equiv& P(U^{,i}U^{,j}) \,, \qquad  P_2 \equiv
P_2^{ii}=\Phi_2
-{1 \over 2}U^2 
\,,\nonumber \\
G_1 &\equiv& P({\dot U}^2)  \,, \qquad  G_2 \equiv P(U {\ddot U}) \,,
\nonumber \\
G_3 &\equiv& -P({\dot U}^{,k} V^k) \,, \qquad  G_4 \equiv
P(V^{i,j}V^{j,i}) \,,\nonumber \\
G_5 &\equiv& -P({\dot V}^k U^{,k}) \,, \qquad  G_6 \equiv P(U^{,ij}
\Phi_1^{ij}) \,,\nonumber \\
G_7^i &\equiv& P(U^{,k}V^{k,i}) + {3 \over 4} P(U^{,i}\dot U ) \,,
\qquad  H \equiv P(U^{,ij} P_2^{ij}) \,.
\label{potentiallist}
\end{eqnarray}

The multipole moments that appear in 2.5PN and 3.5PN terms are defined
by
\begin{eqnarray}
P^\mu &\equiv&  \int_{\cal M} \tau^{\mu 0} d^3x \,, 
\nonumber \\
{\cal I}^Q &\equiv& \int_{\cal M} \tau^{00} x^Q d^3x \,,
\nonumber \\
{\cal J}^{iQ} &\equiv& \epsilon^{iab}\int_{\cal M} \tau^{0b}
x^{aQ} d^3x \,,
\nonumber \\
{\cal M}^{ijQ}  &\equiv& \int_{\cal M} \tau^{ij} 
{x}^{Q} d^3 x \;.
\label{definemoments}
\end{eqnarray}
To the order needed for our purposes, 
\begin{eqnarray}
\tau^{00} &=& \sigma - \sigma^{ii} + 4\sigma U_\sigma
	- {7 \over 8\pi} \nabla U_\sigma^2 \,, 
\nonumber \\
\tau^{0i}&=& \sigma^i + 4\sigma^i U_\sigma
	+{2 \over \pi} U_\sigma^{,j} V_\sigma^{[j,i]}
	+{3 \over 4\pi} {\dot U}_\sigma U_\sigma^{,i} \,,
\nonumber \\
\tau^{ij} &=& \sigma^{ij}
	+{1 \over 4\pi} \left ( U_\sigma^{,i}U_\sigma^{,j} -
	{1 \over 2} \delta^{ij} \nabla U_\sigma^2 \right ) \,.
\label{tauPN}
\end{eqnarray}

\section{Treatment of ``spherical'' bodies in PN expansions}
\label{sec:spherical}

We define our bodies to be spherical in a suitably chosen comoving
frame.  For a given body $A$, we choose a frame that momentarily has
the same coordinate velocity $\bf v$ relative to the global PN frame
as body $A$.  Also, in the limit $m_A \to 0$, the frame is locally
Lorentzian with its origin at ${\bf x}_A$, {\it i.e.} the frame is a
local, freely falling frame in the field of the other body.
In that frame, with coordinates $({\hat t},{\hat x}^i)$, the conserved
baryon density
distribution of body $A$ is taken to be spherically symmetric and static,
{\it i.e.} $\rho^* ({\hat t},{\hat {\bf x}}^i) 
\equiv \rho_S (|{\hat {\bf x}}|)$.
We define the baryonic mass, center of mass and velocity of body $A$
according to Eqs. (\ref{rhostardefinitions}).

Our goal is to calculate PN potentials 
and to integrate them over one of the bodies
using the fact that $\rho^*$ is spherical and static in the local ``hatted''
coordinates.  The general form of the integrals to be evaluated is
$\int \int \rho^*(t,{\bf x}) \rho^*(t,{\bf x}^\prime) 
f({\bf x},{\bf x}^\prime) d^3x d^3x^\prime$.
First we note that the quantity $\rho^* (t,{\bf x}) d^3x
=\rho (t,{\bf x}) u^\mu \sqrt{(-g)} d^3\Sigma_\mu$ is a
scalar, {\it i.e.} is the same at a given event in any coordinate
system.   Thus we only need a transformation of the integration
variables $x$ and  $x^\prime$ in $f({\bf x},{\bf x}^\prime)$ to 
the hatted coordinates.
Notice that $x$ and $x^\prime$ are taken at the same global coordinate time
$t$, but will not necessarily
be at the same local coordinate time $\hat t$.

Consider the event $\cal P$ inside the fluid at ${\vec x}^\prime$ and
the center of mass event $\cal Q$ at ${\vec x}_A$, both at the same
time $t$ as the field point $\vec x$ (see Figure 1).  The field point
could be within body $A$ itself, in free space, 
or in the other body.  Define ${\bar x}
= {\vec x}^\prime - {\vec x}_A = {\bar x}^j {\vec e}_j$, where ${\vec
e}_\mu$ are the basis vectors of our global PN frame; ${\bar x}$ is
purely spatial in this frame.  In the local comoving frame 
\begin{equation}
{\bar x} = {\bar x}^j {\vec e}_j = 
x^{\hat \mu} {\vec e}_{\hat \mu} = 
{\hat t} {\vec e}_{\hat 0} + {\hat x}^j {\vec e}_{\hat j} \,.
\label{transform1}
\end{equation}
Let the transformation between the two basis vectors
take the general form 
\begin{equation}
{\vec e}_{\mu} = ( \Lambda^{\hat \alpha}_\mu 
+ B^{\hat \alpha}_{\mu\nu} {\bar x}^\nu) {\vec e}_{\hat \alpha} \,,
\label{transform2}
\end{equation}
where ${\bar x}^\nu = x^\nu - x_A^\nu $, and 
$B^{\hat \alpha}_{\mu\nu}$ is symmetric on the lower indices.  
The coefficients 
$\Lambda^{\hat \alpha}_\mu$ correspond to boosts and coordinate
re-scalings, while the
coefficients
$B^{\hat \alpha}_{\mu\nu}$ correspond to making the frame freely
falling.   Substituting Eq. (\ref{transform2}) into
(\ref{transform1})
we obtain
\begin{eqnarray}
\hat t &=&  \Lambda^{\hat 0}_j {\bar x}^j + B^{\hat 0}_{jk} {\bar x}^j
{\bar x}^k \,,
\nonumber \\
{\hat x}^j &=& \Lambda^{\hat j}_k {\bar x}^k + B^{\hat j}_{kl} {\bar
x}^k {\bar x}^l \,.
\label{transform3}
\end{eqnarray}
Notice that $\cal P$ is not simultaneous with $\cal Q$ in hatted
coordinates, and that the time difference $\hat t$ depends on ${\bar
x}^j$.  However, since we assume that $\rho^* ({\hat x}^j, {\hat t})$
is time-independent, this  variation of $\hat t$ with integration
point is irrelevant.  

Writing $\Lambda^{\hat \alpha}_\mu = \delta^\alpha_\mu 
+ {\tilde \Lambda}^{\hat \alpha}_\mu$, it is straightforward to show
that, in a PN expansion, the various coefficients in 
Eq. (\ref{transform2})
have the leading orders 
${\tilde \Lambda}^{\hat 0}_0 \sim {\tilde \Lambda}^{\hat j}_k \sim \epsilon$,
${\tilde \Lambda}^{\hat j}_0 \sim {\tilde \Lambda}^{\hat 0}_j \sim \epsilon
^{1/2}$,
$B^{\hat 0}_{0j} \sim B^{\hat j}_{00} \sim B^{\hat j}_{kl} \sim \epsilon$,
$B^{\hat j}_{0k} \sim B^{\hat 0}_{jk} \sim B^{\hat 0}_{00} \sim \epsilon
^{3/2}$.  The transformation then takes the form 
${\hat x}^j = {\bar x}^j +  {\tilde \Lambda}^{\hat j}_k {\bar x}^k + B^{\hat
j}_{kl} {\bar x}^k {\bar x}^l$, where the corrections are of leading
order $\epsilon$.  Inverting the transformation iteratively, we
obtain
\begin{equation}
x^i = x_A^i + {\hat x}^j \{\delta^i_j + \epsilon ( A^i_j + B^i_{jk}
        {\hat x}^k)
        + \epsilon^2 ( C^i_j  + D^i_{jk} {\hat x}^k
	+ E^i_{jkl} {\hat x}^k {\hat x}^l )
        + O(\epsilon^3) \} \,,
\label{flattening2}
\end{equation}
where the coefficients are functions of the ${\tilde \Lambda}^{\hat
j}_k$ and $B^{\hat j}_{kl}$.  To the PN order at which we are working,
their explicit forms are not needed.  
Notice that, in terms of the scale $s \sim |{\hat x}|$, the flattening
correction terms in Eq. (\ref{flattening2}) have the general form
$\epsilon^n (\alpha + \beta s + \dots + \gamma s^n ) \sim 
\epsilon^n (1+s)^n$.

In the double integral of $f({\bf x},{\bf x}^\prime)$, where the dependence is
generally on the difference ${\bf x}-{\bf x}^\prime$, there are two cases to
consider, one where ${\bf x}$ and ${\bf x}^\prime$ are in different bodies, and
the other where they are in the same body.  In the former case, we
substitute Eq. (\ref{flattening2}) for 
both ${\bf x}^i$ and ${\bf x}^\prime$, and
expand about the quantity ${\bf x}_A-{\bf x}_B$ in powers of $s$
using Eq. (\ref{multipoleexpand}),
convert the quantities
$\rho^*d^3x$ to the hatted coordinates, and then integrate over the
spherical density distributions, keeping only terms of $O(s^0)$.  The
first term in the expansion of Eq. (\ref{flattening2}) produces the normal
multipole expansion of the potential, while the remain terms are
relativistic flattening corrections.
In the case where both ${\bf x}$ and ${\bf x}^\prime$ are in the same body, we
have 
\begin{eqnarray}
(x-x^\prime)^i &=& ({\hat x}-{\hat x}^\prime )^j \{ \delta^i_j +
	\epsilon [ A^i_j + B^i_{jk} ({\hat x}+{\hat x}^\prime )^k ]
\nonumber \\
&& 
+ \epsilon^2 [C^i_j + D^i_{jk} ({\hat x}+{\hat x}^\prime )^k
+ E^i_{jkl} ({\hat x}^k{\hat x}^l + {\hat x}^k{\hat x}^{l\prime}
+ {\hat x}^{k\prime}{\hat x}^{l\prime} ) ] + O(\epsilon^3) \} \,.
\label{flattening4}
\end{eqnarray}
In this case, the corrections come only from relativistic flattening, and
also have the general form $\epsilon^n (1+s)^n$.

\section{Evaluation of nonlinear 2PN potentials for two-body systems}
\label{sec:nonlinearpotentials}

\subsection{Triangle potentials}

The potential $P_2^{ij} = P(U^{,i}U^{,j})$ represents a new kind of
potential that first appears at 2PN order.  Unlike the Newtonian
potential $U$, whose fundamental ingredient $1/|{\bf x}-{\bf
x}^\prime|$ depends on the field point and the source
point, the fundamental ingredient of $P_2^{ij}$ depends on the field
point and on {\it two} source points, hence the name ``triangle''
potential.  The potentials $G_i$, $i=1 \dots 6$ and $G_7^i$ are also triangle
potentials.  To see this, we write $P_2^{ij}$ in the form
\begin{eqnarray}
P_2^{ij} &=& {1 \over 4\pi} \int_{\cal M} 
	{d^3x^\prime \over {|{\bf x}-{\bf x}^\prime|}}
	\int \rho^{*\prime\prime} d^3x^{\prime\prime} 
	{{(x^\prime-x^{\prime\prime})^i}
	\over {|{\bf x}^\prime-{\bf x}^{\prime\prime}|^3}}
	\int \rho^{*\prime\prime\prime} d^3x^{\prime\prime\prime}
	{{(x^\prime-x^{\prime\prime\prime})^j}
        \over {|{\bf x}^\prime-{\bf x}^{\prime\prime\prime}|^3}}
\nonumber \\
&=& \sum_{A,B} \int_A \rho_A^* \nabla_A^i d^3x_A
	\int_B \rho_B^* \nabla_B^j d^3x_B
	{1 \over 4\pi} \int_{\cal M} {d^3x^\prime \over
	{|{\bf x}-{\bf x}^\prime|
	|{\bf x}_A-{\bf x}^\prime|
	|{\bf x}_B-{\bf x}^\prime| }} 
\nonumber \\
&=& \sum_{A,B} \int_A \int_B \rho_A^*  \rho_B^*
          d^3x_A d^3x_B \nabla_A^i \nabla_B^j {\cal G}(xAB)
\,,
\label{P2def}
\end{eqnarray}
where 
\begin{eqnarray}
{\cal G}(xAB) &\equiv& - \ln \Delta(xAB) + 1 \,,
\nonumber \\
\Delta(xAB) &\equiv& |{\bf x}-{\bf x}_A|+
        |{\bf x}-{\bf x}_B|+
        |{\bf x}_A-{\bf x}_B| \,.
\label{GDeltadef}
\end{eqnarray}
The triangle function ${\cal G}(ABC)$ satisfies the equations
\begin{eqnarray}
\nabla_A^2 {\cal G}(ABC) &=& - 1/(r_{AB}r_{AC}) \,,
\nonumber \\
\nabla_A^i \nabla_B^i {\cal G}(ABC) &=& {1 \over 2} 
	\left [ {1 \over r_{AB}} \left ( {1 \over r_{AC}} +
	{1 \over r_{BC}} \right ) - {1 \over r_{AC}}{1 \over r_{BC}}
	\right ] \,,
\label{Gequations}
\end{eqnarray}
together with the obvious results obtained by interchange of indices.
Specific gradients of ${\cal G}(xAB)$ have the form
\begin{eqnarray}
\nabla_B^j {\cal G}(xAB) &=& {1 \over \Delta(xAB)} 
	({\hat y}_B + n_{AB})^j \,,
\nonumber \\
\nabla_A^i \nabla_B^j {\cal G}(xAB) &=& {1 \over \Delta(xAB)^2}
	({\hat y}_A - n_{AB} )^i ({\hat y}_B + n_{AB})^j	
	+{1 \over {r_{AB}\Delta(xAB)}}
	(\delta^{ij} - n_{AB}^i n_{AB}^j ) \,,
\nonumber \\
\nabla_B^i \nabla_B^j {\cal G}(xAB) &=& {1 \over \Delta(xAB)^2}
	 ({\hat y}_B + n_{AB} )^i ({\hat y}_B + n_{AB})^j 
	-{1 \over {r_{AB}\Delta(xAB)}}
        (\delta^{ij} - n_{AB}^i n_{AB}^j )
\nonumber \\
&&
	-{1 \over {y_{B}\Delta(xAB)}}
        (\delta^{ij} - {\hat y}_{B}^i {\hat y}_{B}^j ) \,,
\label{Ggradients}
\end{eqnarray}
where $y_A^i  \equiv (x-x_A)^i$, $y_A \equiv |{\bf y}_A|$, and 
${\hat y}_A^i =y_A^i/y_A$.
With these definitions, the other triangle potentials may be written
\begin{mathletters}
\begin{eqnarray}
G_1 &=& \sum_{A,B} \int_A \int_B \rho_A^*  \rho_B^* d^3x_A d^3x_B 
v_A^iv_B^j \nabla_A^i \nabla_B^j {\cal G}(xAB) \,,
\\
G_2 &=& \sum_{A,B} \int_A \int_B \rho_A^*  \rho_B^* d^3x_A d^3x_B
( a_B^i \nabla_B^i + v_B^iv_B^j \nabla_B^i \nabla_B^j ) {\cal G}(xAB)
\,,
\\
G_3 &=& \sum_{A,B} \int_A \int_B \rho_A^*  \rho_B^* d^3x_A d^3x_B
v_A^iv_B^j \nabla_B^i \nabla_B^j {\cal G}(xAB) \,,
\\
G_4 &=& \sum_{A,B} \int_A \int_B \rho_A^*  \rho_B^* d^3x_A d^3x_B
v_A^iv_B^j \nabla_A^j \nabla_B^i {\cal G}(xAB) \,,
\\
G_5  &=& \sum_{A,B} \int_A \int_B \rho_A^*  \rho_B^* d^3x_A d^3x_B
(a_A^i \nabla_B^i + v_A^iv_A^j \nabla_A^j \nabla_B^i)  {\cal G}(xAB)
\,,
\\
G_6  &=& \sum_{A,B} \int_A \int_B \rho_A^*  \rho_B^* d^3x_A d^3x_B
v_A^iv_A^j \nabla_B^i \nabla_B^j {\cal G}(xAB) \,,
\\
G_7^i &=& \sum_{A,B} \int_A \int_B \rho_A^*  \rho_B^* d^3x_A d^3x_B
(v_A^j - {3 \over 4} v_B^j )\nabla_A^i \nabla_B^j  {\cal G}(xAB) \,.
\end{eqnarray}
\label{PGdefs}
\end{mathletters}

\subsection{The triangle potential $P_2^{ij}$ }

For a two-body system, we may write $P_2^{ij} = 
P_{2(11)}^{ij} +  2P_{2(12)}^{(ij)} +  P_{2(22)}^{ij}$,
where the subscripts in parentheses denote the contributions from the
bodies.  
The contributions $P_{2(11)}^{ij}$ and $P_{2(22)}^{ij}$ from 
a single body with a spherically symmetric mass distribution
$\rho^* = \rho(r)$ 
can be derived in a simple manner.  For
spherical symmetry, the equation for $P_2^{ij}$ takes the form
$\nabla^2 P_2^{ij} = -{n}^{ij} (U^\prime)^2$, where
$U^\prime = dU/dr = -m(r)/r^2$, with $dm(r)/dr = 4\pi\rho(r) r^2$.
Writing 
\begin{equation}
P_2^{ij} \equiv (1/3)\delta^{ij} P_2 + {n}^{<ij>}PT_2 \,,
\label{P2split}
\end{equation}
where ${n}^{<ij>} = {n}^{ij} - \delta^{ij}/3$,
it is easy to show that the trace $P_2$ and the traceless part $PT_2$
satisfy the equations $\nabla^2 P_2 = - (U^\prime)^2$, and 
$\nabla^2 PT_2 -6PT_2/r^2 = -(U^\prime)^2$.  Demanding that the
solutions be regular at the origin and vanish at infinity yields 
\begin{eqnarray}
P_2 &=& - {m(r)^2 \over 2r^2} + {8\pi \over r} \int_0^r
	\rho(r^\prime)m(r^\prime)r^\prime dr^\prime
	+ 4\pi \int_r^R \rho(r^\prime)m(r^\prime)dr^\prime
\,,
\nonumber \\
PT_2 &=&  {m(r)^2 \over 4r^2} - {8\pi \over 5r^3}\int_0^r
        \rho(r^\prime)m(r^\prime)r^{\prime 3} dr^\prime 
	+ {{2\pi r^2} \over 5} \int_r^R {\rho(r^\prime)m(r^\prime)
	\over r^{\prime 2}} dr^\prime \,,
\label{P2PT2}
\end{eqnarray}
where $R$ is the radius of the body.  Inside the body, $P_2^{ij} \sim
(m/s)^2$.  Outside the body, $P_2^{ij} = (m/2r)^2 ({n}^{ij}
- \delta^{ij})$, neglecting internal-structure terms of $O(s^{-1})$
and $O(s^{+1})$.  

For the term $P_{2(12)}^{(ij)}$, and for a field point between the two
bodies, we combine the definition (\ref{P2def}) with the
appropriate gradient of $\cal G$ from Eq. (\ref{Ggradients}), and show
that only point mass terms contribute.  Then, for an interbody field
point, we have
\begin{eqnarray}
P_2^{ij} &=& {m_1^2 \over 4y_1^2} ({\hat y}_1^{ij} - \delta^{ij} )
       + {m_2^2 \over 4y_2^2} ({\hat y}_2^{ij} - \delta^{ij} )
\nonumber \\
        && + {2m_1m_2 \over \Delta(x12)}
        \left ( {{({\hat y}_1 - {n})^{(i} ({\hat y}_2 + {\hat
	n})^{j)}}
        \over \Delta(x12)}
        + {{\delta^{ij} - {n}^{ij} } \over r} \right )
        \,,
\label{P2twobody}
\end{eqnarray}
where ${n}^i = x_{12}^i/r$ and $\Delta(x12)
= y_1 +y_2 +r$, and where self-energy terms of $O(s^{-1})$ 
and $O(s^{+1})$ have been
dropped from $P_2^{jk}$. 

\subsection{Evaluation of triangle terms for two-body systems}

We then take either spatial derivatives (e.g. $G_1^{,i}$) 
or time derivatives (e.g. ${\dot G}_7^{i}$) of the triangle
potentials, and in some cases multiply them by other factors (e.g.
$v^jv^k P_2^{ij,k}$ or $U^{,j}P_2^{ij}$), 
and integrate them over the density of body \#1.  Because
we are dealing with a two-body system, then either two or all three of the
points $x$, $A$ and $B$ in ${\cal G}(xAB)$ will reside in the same body,
in other words, in Eqs. (\ref{PGdefs}), we encounter the possibilities 
${\cal G}(111)$,
${\cal G}(112)$,
${\cal G}(121)$, and
${\cal G}(122)$.  
The quantity ${\cal G}(111)$ and its derivatives are purely internal
to one body and can be treated fairly simply.  Notice that a single
derivative of ${\cal G}(111)$
is of the general form $s^j/s^2$, while a double derivative is of
the form $s^{ij}/s^4$, and a general derivative is of the form $s^Q/s^{2q}$,
where $s^Q = s^{i_1 \dots i_q}$.   Since we must be left with one
spatial index $i$, such purely internal terms have odd parity, and
must integrate to zero. 
For the mixed-body cases,
we must expand the functions $\cal G$ about the centers of mass of each
body, sort the terms in powers of the scale $s$ for each body, and retain only
final contributions 
of order $s^0$.  This will be aided by a general expansion of
the function ${\cal G}(ABC)$ in powers of
$r_{AB}/r_{AC}$, where points $A$ and $B$ are assumed to lie inside
one body, and point $C$ is inside the other body, so that $r_{AB} \sim s
\ll r_{AC}$.  
Straightforward
methods lead to the expansion
\begin{eqnarray}
{\cal G}(ABC) &=& -\ln r_{AC} + 1 - \ln 2  
+ {1 \over 2} \sum_{m=0}^\infty
	{{(- r_{AB})^{m+1}} \over {(m+1)!}}
\nonumber \\
&& 
\times	\biggl \{ ({n}_{AB})^M \nabla_A^M \left ( {1 \over r_{AC}} \right)
	+ {r_{AC} \over m+1} ({n}_{AB})^{M+1} 
	\nabla_A^{M+1} \left ( {1 \over r_{AC}} \right ) \biggr \}
\,.
\label{calGsum}
\end{eqnarray}
Note that each term in the expansion is of order $r_{AB}^{m+1} \sim s^{m+1}$,
and depends on gradients of $1/r_{AC}$.  Since $A$ and $C$ are in different
bodies, $\nabla^2 r_{AC}^{-1} =0$; this fact will be important in the
considerations to follow.
The first term in the braces produces a contribution
of order $s^{m+1}$ but of parity (number
of $n_{AB}^i$ vectors ) $(-1)^m$; no matter how many gradients
are taken with respect to any of the variables, this relationship will
be unchanged.  Hence a term of order $s^0$ will have odd parity, a
term of order $s^{-1}$ or $s^{+1}$ will have even parity, and so on.
(Because of the additional scalar factor $r_{AB}$, enough gradients with
respect to $A$ or $B$ will
generate terms of negative powers in $s$.)
The second term in the braces has parity $(-1)^{m+1}$ and order
$s^{m+1}$, a relationship again preserved under any gradients.
Furthermore, gradients of this term will yield terms either completely
independent of $s$ or of positive powers in $s$; no negative powers of
$s$ or terms proportional only to the unit vector ${n}_{AB}^i$ 
can be produced by this term.

Armed with these characteristics of $\cal G$,
consider as an example, the term $G_1^{,i}$ in the equation of motion.
Taking the gradient of Eq. (\ref{PGdefs}a) with respect to $x$, 
pulling out the velocities,
integrating over body \#1,
and considering all the possible cases for
$A$ and $B$, we first 
find no contribution from $\nabla_x^i \nabla_A^j \nabla_B^k
{\cal G}(111) \sim s^is^js^k/s^6$, by symmetry.
Considering the other cases, 
${\cal G}(112)$,
${\cal G}(121)$, and
${\cal G}(122)$, we find that the only $s^0$ contributions from the
gradients of the 
expansion (\ref{calGsum})
are either odd parity (from the first term ), 
and thus vanish on integrating over
body \#1 or \#2, or are independent of $s$ (from the second term), 
and yield the desired
``point mass'' result:
$G_1^{,i} \to m_1m_2 [n^i(4\nva\nvb-\v1v2) -2v_1^i\nvb
-v_2^i\nva]/2r^3-m_2^2 [2n^i(\nvb)^2-n^iv_2^2-v_2^i\nvb]/2r^3$.

Consider as a 
less straightforward example, the term in $G_2^{,i}$ that depends on
the acceleration.  Inserting the Newtonian equation of motion
$a_B^i=U_B^{,i}$,
we must evaluate the term
\begin{equation}
{1 \over m_1} \int_1 \rho^* d^3x 
  	\sum_{A,B} \int_A \int_B \rho_A^*  \rho_B^*
          d^3x_A d^3x_B (U_{B,int}^{,j} + U_{B,ext}^{,j} )
	\nabla_x^i \nabla_B^j {\cal G}(xAB) \,,
\label{G2}
\end{equation}
where we have split $U$ into a contribution from within body $B$
itself (``int'') and from the other body (``ext'').  The case $A=B=1$ has a
purely internal term from $U_{1,int}^{,j}\nabla_x^i \nabla_B^j{\cal G}(111)$, 
which vanishes by
symmetry, and a term $U_{1,ext}^{,j} \nabla_x^i \nabla_B^j {\cal G}(111) 
\sim m_2\sum_q ({\bar x}^Q \nabla^Q \nabla^j r^{-1}/q!) s^is^j/s^4$, where we
expand the external potential about the center of mass of body \#1, and
where $r=r_{12}$.  
Only the $q=2$ term
contributes at overall order $s^0$, leading to an integral of the 
schematic form $\int_1 ({\bar x}^{kl}
s^is^j/s^4) \nabla^{klj} r^{-1} \to
(\delta^{kl}\delta^{ij}+\delta^{ki}\delta^{jl}+\delta^{kj}\delta^{il})
\nabla^{klj} r^{-1} \propto \nabla^2 \nabla^i r^{-1} = 0$.  
Similarly, combining $U_{B,int}^{,j} \sim s^j/s^3$ with the expansion of 
$\nabla_x^i \nabla_B^j {\cal G}(112)$ produces a potentially $s^0$ term only
from an order $s^2$ term from the derivatives of $\cal G$; but such terms
are necessarily accompanied by several (three or more) 
gradients of $r^{-1}$; because only
a single index $i$ remains at the end, two of the gradients are always
contracted
into $\nabla^2$, which 
vanishes when acting on $r^{-1}$.  All the possible combinations of $A$ and
$B$ in Eq. (\ref{G2})
yield equivalent results.  The final answer contains only ``point''
mass terms, and is equivalent to combining the 
completely
$s$-independent terms from the
derivatives of ${\cal G}$ with only the ``external'' potential terms arising
from any acceleration.  The same approach holds for the terms $G_5^{,i}$,
${\dot G}_7^i$, and $U^{,j}P_2^{ij}$.

The vanishing of many potential contributions at order $s^0$ depends
critically on the fact that the terms ultimately depend on 
the factor $\nabla^{klj} r^{-1}$, and that two of the
indices are contracted, 
since only one index is allowed to be free.  However, at 3PN order,
this is no longer the case.  A simple example is provided by a  3PN term
proportional to $v^jv^kU^{,i}P_2^{jk}$.  Integrating over body \#1, 
the combination
of $U_{ext}^{,i}$ with $P_{2(11)}^{jk}$ gives a contribution to the equation
of motion
\begin{eqnarray}
{1 \over m_1} v_1^j && v_1^k \int_1 \rho^* d^3{\bar x} \,m_2\sum_q 
	{{\bar x}^Q \over q!} \nabla^Q \nabla^j (r^{-1}) 
	\left ( {1 \over 3} P_2({\bar r}) \delta^{jk}
	+ PT_2({\bar r}) n^{<jk>} \right )
\nonumber \\
&& = {1 \over 15} \left \{ \int_0^R \rho^* {\bar r}^2 PT_2({\bar r}) d^3{\bar x}
	/m_1^3 \right \}
	m_1^2 m_2 v_1^jv_1^k \nabla^{ijk} (r^{-1})  \,,
\end{eqnarray}
where the 
quantity in braces is dimensionless, scales as $s^0$ for fixed $m_1$,
but depends on the
internal structure of body \#1.  Contributions like this appear everywhere
at 3PN order; whether they survive in the final equation of motion, or what
their ultimate interpretation is, will be the subject of future work.

\subsection{The quadrangle potential $H$}

The potential $H = P(U^{,ij}P_2^{ij})$ 
is an example of a more complicated ``quadrangle''
potential whose fundamental ingredient depends on the field point and three
source points.  
To see this, we write
\begin{eqnarray}
H &=& {1 \over 4\pi} \int_{\cal M}
	{d^3x^\prime \over {|{\bf x}-{\bf x}^\prime|}}
	U^{,ij}(x^\prime) P_2^{ij}(x^\prime)
\nonumber \\
&=& {1 \over 4\pi} \int_{\cal M}
        {d^3x^\prime \over {|{\bf x}-{\bf x}^\prime|}}
	U^{,ij}(x^\prime)
	{1 \over 4\pi} \int_{\cal M}
	{d^3x^{\prime\prime} \over 
	{|{\bf x}^\prime-{\bf x}^{\prime\prime}|}}
	U^{,i}(x^{\prime\prime}) U^{,j}(x^{\prime\prime})
\nonumber \\
&=& \sum_{ABC} 
	\int_A \rho_A^* \nabla_A^i \nabla_A^j d^3x_A
	\int_B \rho_B^* \nabla_B^i d^3x_B
	\int_C \rho_C^* \nabla_C^j d^3x_C
	 \, {\cal H}(xA;BC) \,,
\label{Hdef}
\end{eqnarray}
where the function ${\cal H}$ of four field points is defined by
\begin{equation}
{\cal H}(AB;CD) \equiv {1 \over (4\pi)^2} \int_{\cal M}\int_{\cal M}
	{{d^3x^\prime d^3x^{\prime\prime}} \over 
	{|{\bf x}_A-{\bf x}^\prime|
	|{\bf x}_B-{\bf x}^\prime|
	|{\bf x}^\prime-{\bf x}^{\prime\prime}|
	|{\bf x}_C-{\bf x}^{\prime\prime}|
	|{\bf x}_D-{\bf x}^{\prime\prime}|}} \,,
\label{calHdef}
\end{equation}
with the properties
\begin{eqnarray}
\nabla_A^2 {\cal H}(AB;CD) &=& -(1/r_{AB}) {\cal G}(ACD) \,,
\nonumber \\
\nabla_C^2 {\cal H}(AB;CD) &=& -(1/r_{CD}) {\cal G}(ABC) \,,
\nonumber \\
\nabla_A^i \nabla_B^i {\cal H}(AB;CD) &=& (1/2r_{AB})[{\cal G}(ACD) +
{\cal G}(BCD)] - {\cal J}(ABCD)/2 \,,
\label{calHprops}
\end{eqnarray}
where
\begin{equation}
{\cal J}(ABCD) \equiv {1 \over 4\pi} \int_{\cal M} 
	{d^3x^\prime \over
	{|{\bf x}_A-{\bf x}^\prime||{\bf x}_B-{\bf x}^\prime|
	|{\bf x}_C-{\bf x}^\prime||{\bf x}_D-{\bf x}^\prime|}} \,.
\end{equation}
Unfortunately, we have been unable to find a closed-form solution for
$\cal H$ similar to that for $\cal G$, nor a useful expansion in the
case where some of the distances between points are small compared to
the others.  

Instead, we make use of the first form of $H$ given in
Eq. (\ref{Hdef}).  
We integrate $H^{,i}$ over the density of body \#1
and substitute $U^{,k} = U_1^{,k}+U_2^{,k}$
and $ P_2^{jk} =  P_{2(11)}^{jk} +  2P_{2(12)}^{(jk)} +  P_{2(22)}^{jk}$.
The result can be put into the form
\begin{equation}
\int_1 \rho^* H^{,i} d^3x = {1 \over 4\pi} \int_{\cal M}
        d^3x^\prime U_1^{,i} (x^\prime)
	(U_1^{,jk} (x^\prime)+U_2^{,jk} (x^\prime)) 
	( P_{2(11)}^{jk} +  2P_{2(12)}^{(jk)} +  P_{2(22)}^{jk}) \,.
\label{Hintegral}
\end{equation}
We wish to verify that no contributions of order $s^0$ (other
than normal point mass terms) arise in this integral.  
To see this, we split the integral over $\cal M$ into an integral over body
\#1 out to a radius $\sim s_1$, a similar integral over body \#2 to a
radius $\sim s_2$, and an integral over the rest of $\cal M$.  
In the latter integral, we may use solutions external to each spherical
body:
$U = m_1/y_1 + m_2/y_2$ and $P_2^{ij}$ from Eq. (\ref{P2twobody}). 
If carried out over all of $\cal M$ using these
functions, the integral would diverge at the locations of the two bodies
\cite{note1}.  

Consider now the integral over a region of volume $s_1^3$ surrounding 
body \#1.  In the neighborhood of
\#1, the product $U_1^{,i} U_1^{,jk}$ behaves as $s^is^js^k/s^8$;
multiplying by the volume, we have a term of
odd parity and $O(s_1^{-2})$.  The term $P_{2(11)}^{jk}$ is even parity, so the
combination integrates to zero.  The term $P_{2(12)}^{(jk)}$
must be expanded about \#1 using Eq. (\ref{calGsum}); the expansion begins at
$O(s^0)$ with a constant term and an odd parity term proportional to
$s^i/s$; then at $O(s)$ with a term proportional to $s$ (even parity) and
one proportional to $s^i$ (odd parity); then at $O(s^2)$ with terms
proportional to $ss^i$ and $s^is^j$, and so on.
All even parity terms integrate to zero when multiplied by
$s^is^js^k/s^8$.  The odd-parity $O(s^0)$  and $O(s)$ terms lead 
to non-zero integrals of
order $s_1^{-2}$ and $s_1^{-1}$, which we discard. The odd-parity
contribution of order $s^2$, is accompanied by a coefficient proportional
to
$\nabla^{jkl} (1/r)$.  Integrating over the sphere then results in a term
of order $s^0$ but proportional to $\nabla^2 (1/r)$, which vanishes.  
Finally expanding the term $P_{2(22)}^{(jk)}$ about the location of
body \#1 gives only terms of order $s^m$ and parity $(-1)^m$, hence only a
contribution of order $s_1^{-1}$ survives in the integral.  Applying similar
considerations to the product $U_1^{,i} U_2^{,jk}$, and then repeating the
considerations for the integral over the neighborhood of body \#2 leads to
the conclusion that the contributions are of order $s_A^{-2}$ or $s_A^{-1}$,
or of positive powers,
but that there are no structure-dependent contributions of order $s_A^0$.  

Consequently, we can carry out the integral in Eq. (\ref{Hintegral}) 
over $\cal M$ up
to spheres surrounding each body, and then let the spheres shrink to zero,
discarding terms that blow up as $s_A^{-2}$ or $s_A^{-1}$, and keeping only
finite terms.  We are guaranteed that no 
structure-dependent terms of $O(s^0)$ will appear.
The final result is given in Eq. (\ref{Hfinal}).

\section{Evaluation of multipole moments for two-body systems}
\label{sec:moments}

\subsection{Quadrupole and higher moments}
\label{sec:quadrupole}

Substituting expressions for $\tau^{\mu\nu}$ from Eqs. (\ref{tauPN})
into the definitions of the multipole moments, Eqs. (\ref{definemoments}),
converting from densities $\sigma$, $\sigma^i$ and $\sigma^{ij}$ to
the conserved baryon density $\rho^*$ to the needed PN order  
using Eqs. (\ref{sigmatorhoPN}), 
and integrating, discarding any terms that depend
explicitly on the radius $\cal R$ of the boundary of $\cal M$, we
obtain
\begin{eqnarray}
{\cal I}^{ij} &=& \sum_A m_A x_A^{ij} (1+ {1 \over 2}v_A^2)
	- \sum_{AB} {m_Am_B \over r_{AB}} 
	\left ( {1 \over 2} x_A^{ij} - {7 \over 4}  r_{AB}^2
	\delta^{ij} \right ) \,,
\nonumber \\
{\cal I}^{ijk\dots} &=& \sum_A m_A x_A^{ijk\dots} \,,
\nonumber \\
{\cal J}^{ij} &=& \epsilon^{iab} \sum_A m_A v_A^bx_A^{aj} \,,
\nonumber \\
{\cal J}^{ijk\dots} &=& \epsilon^{iab} \sum_A m_A v_A^bx_A^{ajk\dots} \,,
\nonumber \\
{\cal M}^{ijkl} &=& \sum_A m_A v_A^{ij} x_A^{kl} 
	- {1 \over 2} \sum_{AB} {m_Am_B \over r_{AB}} n_{AB}^{ij} x_A^{kl} 	
	+ {1 \over 12} \sum_{AB} m_Am_Br_{AB} 
	\biggl ( n_{AB}^{ijkl} -n_{AB}^{ij} \delta^{kl}
\nonumber \\
&&
	-n_{AB}^{kl} \delta^{ij} 
	+ n_{AB}^{i(k} \delta^{l)j}
	+ n_{AB}^{j(k} \delta^{l)i} - 2 \delta^{i(k} \delta^{l)j}
	+ 2 \delta^{ij}  \delta^{kl} \biggr ) \,.
\label{PNmoments}
\end{eqnarray}
Note that, although ${\cal I}^{ijk}$ and ${\cal J}^{ij}$ appear in
2.5PN terms, they are purely functions of time, and thus  cancel out of
the relative equation of motion, so they are only needed to lowest
order for use in 3.5PN terms.

Converting to relative coordinates, using the 1PN correct
transformation in the leading term of ${\cal I}^{ij}$, we obtain
\begin{eqnarray}
{\cal I}^{ij} &=& m \eta x^{ij} \left ( 1+ {1 \over 2}(1-3\eta)v^2 -
	{1 \over 2}(1-2\eta){m \over r} \right ) 
	+ {7 \over 2} m^2 \eta r \delta^{ij} \,,
\nonumber \\
{\cal I}^{ijk} &=& - \delta m \eta x^{ijk}  \,,
\nonumber \\
{\cal I}^{ijkl} &=& m \eta (1-3\eta)  x^{ijkl} \,,
\nonumber \\
{\cal J}^{ij} &=& -\delta m \eta L^i x^j \,,
\nonumber \\
{\cal J}^{ijk} &=&  m \eta (1-3\eta)  L^i x^{jk} \,,
\nonumber \\
{\cal M}^{ijkl} &=& m \eta (1-3\eta) \left (v^{ij} - 
	{1 \over 3} {m \over r} n^{ij} \right ) x^{kl} 
\nonumber \\
&&
	- {1 \over 6}  m^2 \eta r 
	\left (  n^{ij} \delta^{kl}
	+n^{kl} \delta^{ij} - n^{i(k} \delta^{l)j}
	- n^{j(k} \delta^{l)i} + 2 \delta^{i(k} \delta^{l)j}
	- 2 \delta^{ij}  \delta^{kl} \right ) \,,
\label{relativemoments}
\end{eqnarray}
where ${\bf L} \equiv {\bf x} \times {\bf v}$ is the orbital angular
momentum per unit mass.
Note that the moments ${\cal I}^{ijklm}$, ${\cal J}^{ijkl}$ and ${\cal
M}^{ijklm}$ are not needed explicitly for the 3.5PN equations of motion, since
they are purely functions of time, and cancel out of the relative
equation.

Time derivatives of the moments may be calculated using the relative equations
of motion in place of ${\ddot x}^i$; 1PN equations must be used in the
leading term in ${\cal I}^{ij}$, while Newtonian equations are
sufficient for the remaining terms.

\subsection{Dipole moment and the system center of mass}
\label{sec:dipole}

For a two-body system, the dipole moment is given by
${\cal I}^i = \int_{\cal M} \tau^{00}x^i d^3x$.  Substituting for
$\tau^{00}$ including 1PN and 2.5PN terms from Eq. (5.9) of Paper I, and
convering from $\sigma$ to $\rho^*$ to the corresponding order from
Eq. (\ref{sigmatorhoPN}), we obtain, to zeroth, PN and 2.5PN order,  
\begin{eqnarray}
{\cal I}^i &=& \sum_A m_A x_A^i \left (1 + {1 \over 2}v_A^2 - {1 \over 2} 
	\sum_B {m_B \over r_{AB}} \right ) 
\nonumber
\\
&& - {2 \over 5} \sum_A m_A x_A^j \left ( 
	\delta^{ij}  \stackrel{(3)}{{\cal I}^{kk}}
	+ 2 \stackrel{(3)}{{\cal I}^{ij}} \right ) \,.
\label{dipolePN}
\end{eqnarray}
Choosing the center
of mass so that ${\cal I}^i =0$ to at least PN order, we see that the
final 2.5PN term is in fact at least 3.5PN order.
However, we must also check the time dependence of
${\cal I}^i$, to see if it remains zero to the order needed.  From the
definition of ${\cal I}^i$ (Paper I, Eq. (4.6)), we have
\begin{equation} 
{\dot {\cal I}}^i = P^i - \oint_{\partial {\cal M}} \tau^{0j}x^i
d^2S_j  \,.
\label{dipole2}
\end{equation}
Using the definition of $\tau^{0j}$ (Paper I, Eqs. (4.4)) and the far-zone
forms of the gravitational potentials (Paper I, Eqs. (5.12), 
with ${\cal I}^i =0$ to 1PN
order), it can be shown that the
surface integral is of 2.5PN order relative to $P^i$.  However, taking
an
additional time derivative and using $\tau^{ij}$ in the far zone gives
\begin{eqnarray}
{\ddot {\cal I}}^i = 
{\dot P}^i &=& - \oint_{\partial {\cal M}} \tau^{ij} d^2S_j  \,,
\nonumber \\
&=& - {\cal I} \left ( {2 \over 15}  \stackrel{(5)}{{\cal I}^{ijj}}
	- {2 \over 3} \epsilon^{qik} \stackrel{(4)}{{\cal J}^{qk}}
	\right ) \,,
\label{dipole3}
\end{eqnarray}
where, to the order needed, $\cal I$ is simply the total baryon mass
$m$ of the system.
Hence integrating with respect time and setting the initial conditions
${\cal I}_0^i =0$ and $P_0^i =0$, we have
\begin{equation}
{\cal I}^i = -m \left ( {2 \over 15}  \stackrel{(3)}{{\cal I}^{ijj}}
        - {2 \over 3} \epsilon^{qik} \stackrel{(2)}{{\cal J}^{qk}}
        \right ) \,.
\label{dipole4}
\end{equation} 
Note that, while this may seem like an anomalous 2.5PN effect on the
system center of mass, it is really a gauge effect.  Because the right
hand side of Eq. (\ref{dipole4}) is a total time derivative, it can be absorbed
into a redefinition of spatial coordinates.  Combining Eqs. (\ref{dipolePN}) 
and (\ref{dipole4}),
and defining $x_1^i = (m_2/m)x^i + \zeta^i$, $x_2^i = -(m_1/m)x^i +
\zeta^i$, 
we can then show that the transformation from $x_A^i$ to relative
coordinates is given by Eqs. (\ref{transform}), which 
were derived directly from the
1PN and 2.5PN equations of motion.

\begin{figure}
\begin{center}
\leavevmode
\psfig{figure=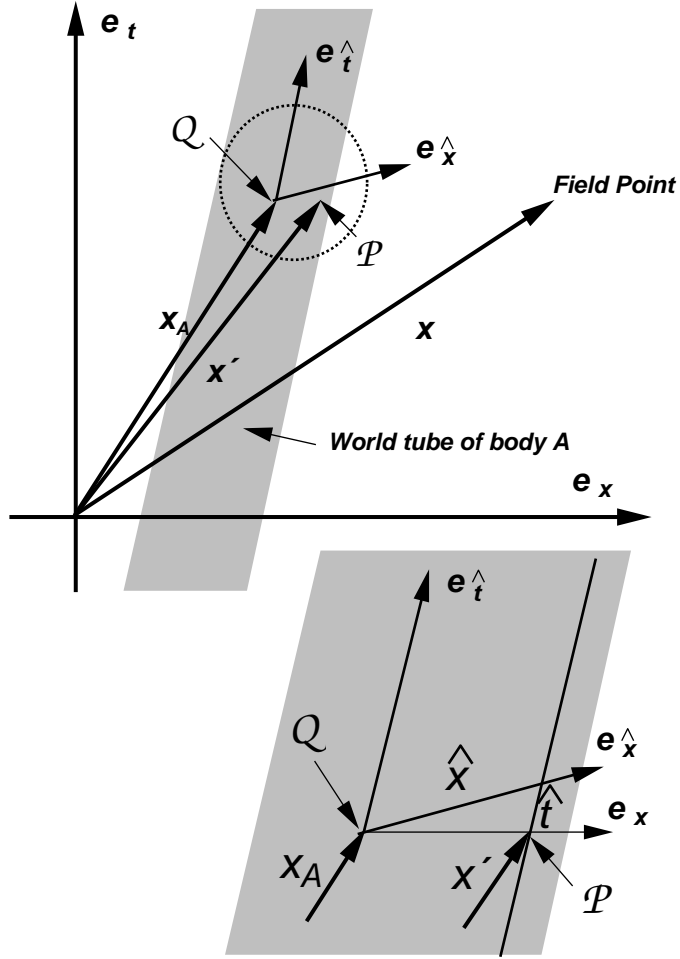,height=5.0in}
\end{center}
\caption{Transformation to the local comoving frame of one body.
Global harmonic coordinates are $t,x^j$; local coordinates are ${\hat
t}, {\hat x}^j$.  In local coordinates, the density of the body is
assumed to be static and spherically symmetric.
}
\label{dire2fig1}
\end{figure}


\begin{references}

\bibitem{3min} C. Cutler, T. A. Apostolatos, L. Bildsten, L. S. Finn,
\'E. E. Flanagan, D. Kennefick, D. M. Markovi\'c, A. Ori, E. Poisson,
G. J. Sussman, and K. S. Thorne, Phys. Rev. Lett. {\bf 70}, 2984
(1993).

\bibitem{patiwill} M. E. Pati and C. M. Will,  
Phys. Rev. D {\bf 62}, 124015
(2000)

\bibitem{DD81} T. Damour and N. Deruelle, Phys. Lett. {\bf 87A},
81 (1981).

\bibitem{damour300} T. Damour, in {\it 300 Years of Gravitation},
edited by S. W. Hawking and W. Israel (Cambridge University Press,
London, 1987), p. 128.

\bibitem{kopeikin85} S. M. Kopeikin, Sov. Astron. {\bf 29}, 516
(1985).

\bibitem{GK86} L. P. Grishchuk and S. M. Kopeikin, In {\it Relativity
in Celestial Mechanics and Astrometry}, edited by J. Kovalevsky and V.
A. Brumberg (Reidel, Dordrecht, 1986), p. 19.

\bibitem{bfp98} L. Blanchet, G. Faye and B. Ponsot, Phys. Rev. D {\bf
58}, 124002 (1998).

\bibitem{futamase01} Y. Itoh, T. Futamase, and H. Asada, 
Phys. Rev. D {\bf 63},
064038 (2001).

\bibitem{MTW} C. W. Misner, K. S. Thorne, and J. A. Wheeler, {\it
Gravitation} (Freeman Publishing Co, San Francisco, 1973).

\bibitem{iyerwill} B. R. Iyer and C. M. Will, Phys. Rev. Lett. {\bf
70}, 113 (1993).

\bibitem{iyerwill2} B. R. Iyer and C. M. Will, Phys. Rev. D {\bf
52}, 6882 (1995).

\bibitem{schafer85} G. Sch\"afer, Ann. Phys. (N.Y.) {\bf 161}, 81
(1985).

\bibitem{schafer86} G. Sch\"afer, Gen. Relativ. Gravit. {\bf 18}, 255
(1986).

\bibitem{jaraschafer97} P. Jaranowski and G. Sch\"afer, Phys. Rev. D
{\bf 55}, 4712 (1997).

\bibitem{jaraschafer98} P. Jaranowski and G. Sch\"afer, Phys. Rev. D
{\bf 57}, 5948 (1998), {\it ibid}. {\bf 57}, 7274 (1998).

\bibitem{jaraschafer99} P. Jaranowski and G. Sch\"afer, Phys. Rev. D
{\bf 60}, 124003 (1999).

\bibitem{djs00} T. Damour, P. Jaranowski and G. Sch\"afer,  Phys. Rev.
D {\bf 62}, 021501 (2000).

\bibitem{bf00} L. Blanchet and G. Faye,  Phys.Lett. {\bf 271A}, 58 (2000).

\bibitem{bf01} L. Blanchet and G. Faye,  Phys. Rev. D
{\bf 63}, 062005 (2001). 

\bibitem{thorne80} K. S. Thorne, Rev. Mod. Phys. {\bf 52}, 299
(1980).

\bibitem{bfhadamard}  L. Blanchet and G. Faye, J. Math. Phys. {\bf 41}, 
7675 (2000).
 
\bibitem{EIH} A. Einstein, L. Infeld, and B. Hoffmann, Ann. Math. {\bf
39},
65 (1938).

\bibitem{futamase00} Y. Itoh, T. Futamase, and H. Asada, 
Phys. Rev. D {\bf
62}, 064002 (2000).

\bibitem{futamase85} T. Futamase, 
Phys. Rev. D {\bf 32}, 2566 (1985); {\it
ibid}. {\bf 36}, 321 (1987).

\bibitem{tegp} C. M. Will, {\it Theory and Experiment in Gravitational
Physics} (Cambridge University Press, Cambridge, 1993).

\bibitem{wagwill} R. V. Wagoner and C. M. Will, Astrophys. J. {\bf
210}, 764 (1976); {\bf 215}, 984 (1977).

\bibitem{bds} L. Blanchet and T. Damour, Ann. Inst. Henri Poincar\'e
A, {\bf
50}, 377 (1989); L. Blanchet and G. Sch\"afer, Mon. Not. R. Astron.
Soc. {\bf 239}, 845 (1989).

\bibitem{junker} W. Junker and G. Sch\"afer, Mon. Not. R. Astron. Soc.
{\bf 254}, 146 (1992).

\bibitem{luc93} L. Blanchet, Phys. Rev. D {\bf 47}, 4392 (1993).

\bibitem{note1} In evaluating integrals of second
derivatives of $U$, it is important
to use the fact that $\nabla^i \nabla^j (1/|{\bf
x} - {\bf x}^\prime |) = 3(x-x^\prime)^{<ij>}/|{\bf
x} - {\bf x}^\prime |^5 - (4\pi/3)\delta^{ij} \delta^3 ({\bf
x} - {\bf x}^\prime) $.
\end{references}
\end{document}